\documentclass[aps,prl,reprint,nofootinbib]{revtex4-1}

\usepackage{amsmath}
\usepackage{amssymb}	
\usepackage{bbold}
\usepackage{bm} 
\usepackage{cancel} 
\usepackage{xcolor}
\usepackage[normalem]{ulem} 
\usepackage{changepage}
\usepackage{float}
\usepackage{mathrsfs}
\usepackage{url}
\usepackage{xfrac}
\usepackage{graphicx}
\usepackage{lipsum} 
\graphicspath{{/}}
\usepackage[font=small,labelfont=bf,justification=raggedright,format=plain,singlelinecheck=false]{caption} 

\newcommand{\comment}[1]{\ignorespaces} 

\renewcommand{\v}[1]{\ensuremath{\mathbf{#1}}} 
\newcommand{\gv}[1]{\ensuremath{\mbox{\boldmath$ #1 $}}} 
\newcommand{\pd}[2]{\frac{\partial #1}{\partial #2}} 
\providecommand{\e}[1]{\ensuremath{\times 10^{#1}}}
\DeclareMathAlphabet{\mathsfit}{\encodingdefault}{\sfdefault}{m}{n}
\SetMathAlphabet{\mathsfit}{bold}{\encodingdefault}{\sfdefault}{bx}{n}
\newcommand{\ms}[1]{\bm{\mathsfit{#1}}} 
\newcommand{\ws}[1]{\mathsf{#1}} 

\makeatletter
\newcommand*\bigcdot{\mathpalette\bigcdot@{.5}}
\newcommand*\bigcdot@[2]{\mathbin{\vcenter{\hbox{\scalebox{#2}{$\m@th#1\bullet$}}}}}
\makeatother

\makeatletter
\newcommand{\raisemath}[1]{\mathpalette{\raisem@th{#1}}}
\newcommand{\raisem@th}[3]{\raisebox{#1}{$#2#3$}}
\makeatother

\usepackage{accents}

\usepackage{stackengine} 

\makeatletter 
\renewcommand*\env@matrix[1][*\c@MaxMatrixCols c]{%
  \hskip -\arraycolsep
  \let\@ifnextchar\new@ifnextchar
  \array{#1}}
\makeatother

\usepackage{pgfplots}
\usepgfplotslibrary{patchplots}
\usepgfplotslibrary{external}
\usepackage{tikz}
\usetikzlibrary{external}
\usetikzlibrary{calc}
\usetikzlibrary{arrows,shapes,backgrounds,arrows.meta,shapes.arrows,automata,quotes,intersections}
\usetikzlibrary{decorations.text}

\usetikzlibrary{decorations.pathreplacing}
\makeatletter
    \let\pgf@decorate@@brace@brace@code@old\pgf@decorate@@brace@brace@code
    \def\pgf@decorate@@brace@brace@code{
        \ifdim\pgfdecoratedremainingdistance<4\pgfdecorationsegmentamplitude
            \pgftransformxscale{\pgfdecoratedremainingdistance/4\pgfdecorationsegmentamplitude}
            \pgfdecoratedremainingdistance=4\pgfdecorationsegmentamplitude
        \fi
        \pgf@decorate@@brace@brace@code@old
    }
\makeatother

\newcommand*\mD{\mathrm{D}}

\newcommand*\mOne{\mathbb{1}}
\newcommand*\mZero{\mathbb{0}}
\newcommand*\vphi{\varphi}
\newcommand*\tphi{\tilde{\phi}}

\newcommand*\mR{\mathbb{R}}

\newcommand*\md{\mathrm{d}}

\newcommand*\n{[\v{n}]}
\newcommand*\nihat{[\v{n}+\v{i}]}
\DeclareMathAlphabet{\mathpzc}{OT1}{pzc}{m}{it}

\newcommand*\tleft{\triangleright}
\newcommand*\tright{\triangleleft}

\usepackage{bbding}
\DeclareFontFamily{U}{MnSymbolC}{}
\DeclareSymbolFont{MnSyC}{U}{MnSymbolC}{m}{n}
\DeclareMathSymbol{\diamondplus}{\mathbin}{MnSyC}{"7C}
\DeclareMathSymbol{\diamonddot}{\mathbin}{MnSyC}{"7E}
\DeclareFontShape{U}{MnSymbolC}{m}{n}{
    <-6>  MnSymbolC5
   <6-7>  MnSymbolC6
   <7-8>  MnSymbolC7
   <8-9>  MnSymbolC8
   <9-10> MnSymbolC9
  <10-12> MnSymbolC10
  <12->   MnSymbolC12}{}

\makeatletter
\newcommand*{\defeq}{\mathrel{\rlap{%
                     \raisebox{0.3ex}{$\m@th\cdot$}}%
                     \raisebox{-0.3ex}{$\m@th\cdot$}}%
                     =}
\newcommand*{\eqdef}{=\mathrel{\rlap{%
                     \raisebox{0.3ex}{$\m@th\cdot$}}%
                     \raisebox{-0.3ex}{$\m@th\cdot$}}%
                     }
\makeatother

\usepackage{mathtools}

\usepackage{stmaryrd}

\usepackage{enumitem}

\usepackage{diagrams}
\usetikzlibrary{cd}

\definecolor{light-gray}{gray}{.85}
\newsavebox{\songboxbox}

\usepackage[most]{tcolorbox}
\newtcbox{\MyBox}[1][red]{on line, size=tight, boxsep=1pt, colframe=#1!50!black, colback=#1!10!white}

\usepackage[framemethod=TikZ]{mdframed}
\mdfsetup{%
    outerlinewidth=1pt,
    innertopmargin=6pt,
    innerbottommargin=6pt,
    skipabove=10pt,
    skipbelow=10pt,
    backgroundcolor=black!10,
    roundcorner=20pt
}

\DeclareFontFamily{U}{matha}{\hyphenchar\font45}
\DeclareFontShape{U}{matha}{m}{n}{
      <5> <6> <7> <8> <9> <10> gen * matha
      <10.95> matha10 <12> <14.4> <17.28> <20.74> <24.88> matha12
      }{}
\makeatletter
\newcommand{\blandor}[1]{\mathbin{\@blandor{#1}}}
\newcommand{\@blandor}[1]{\mathchoice
  {\@@blandor{#1}{\tf@size}}
  {\@@blandor{#1}{\tf@size}}
  {\@@blandor{#1}{\sf@size}}
  {\@@blandor{#1}{\ssf@size}}
}
\newcommand{\@@blandor}[2]{%
    \raisebox{.1ex}{\rotatebox[origin=c]{#1}{%
      \fontsize{#2}{#2}\usefont{U}{matha}{m}{n}\symbol{\string"CE}}}%
}
\makeatother

\usepackage{slashed} 
\newcommand{\cmmnt}[1]{\ignorespaces} 

\usepackage{environ}
\NewEnviron{eqn}{
\begin{align}\begin{split}
\BODY
\end{split}\end{align}
}

\newcommand{\invcircledast}{%
  \mathbin{\vphantom{\circledast}\text{%
    \ooalign{\smash{\blackcircle}\cr
             \hidewidth\smash{\textcolor{white}{$*$}}\hidewidth\cr
            }%
  }}%
}

\newcommand{\blackcircle}{\raisebox{-.4ex}{\scalebox{1.66}{$\bullet$}}}

\begin{document}


\title{Restoring Poincar\'e Symmetry to the Lattice} 



\author{Alexander S. Glasser}
\affiliation{Department of Astrophysical Sciences, Princeton University, Princeton, New Jersey 08540}
\author{Hong Qin}
\affiliation{Department of Astrophysical Sciences, Princeton University, Princeton, New Jersey 08540}


\date{\today}

\begin{abstract}
The following work demonstrates the viability of Poincar\'e symmetry in a discrete universe.  We develop the technology of the \emph{discrete principal Poincar\'e bundle} to describe the pairing of (1) a hypercubic lattice `base manifold' labeled by integer vertices---denoted $\{\v{n}\}=\{(n_t,n_x,n_y,n_z)\}$---with (2) a Poincar\'e structure group.  We develop \emph{lattice 5-vector theory}, which describes a non-unitary representation of the Poincar\'e group whose dynamics and gauge transformations on the lattice closely resemble those of a scalar field in spacetime. We demonstrate that such a theory generates \emph{discrete} dynamics with the complete \emph{infinitesimal} symmetry---and associated invariants---of the Poincar\'e group.  Following our companion paper \cite{glasser_continuous5vectortheory}, we `lift' the Poincar\'e gauge symmetries to act only on vertical matter and solder fields, and recast `spacetime data'---stored in the $\partial_\mu\phi(x)$ kinetic terms of a free scalar field theory---as `matter field data'---stored in the $\phi^\mu\n$ components of the 5-vector field itself.  We gauge 5-vector theory to describe a lattice gauge theory of gravity, and discuss the physical implications of a discrete, Poincar\'e-invariant theory.
\end{abstract}


\maketitle 



\section{Introduction}
It is often assumed that the lattice discretization of field theories necessarily sacrifices the \emph{infinitesimal} symmetries of the Poincar\'e group.  For a hypercubic lattice in (3+1)-dimensional spacetime, for example, it is noted that the 4 translation symmetries are finite---and equal to the lattice spacing; worse, the 3 rotation symmetries of the Lorentz group are seemingly limited to multiples of $\frac{\pi}{2}$; and worse still, the 3 Lorentz boosts seem to be entirely absent.  The forfeiture of these 10 infinitesimal symmetries invalidates the Noether procedure \cite{noether_invariant_1971}, leading to a loss of energy-momentum conservation and apparently thwarting the fundamental validity of any such lattice theory.

In the following work, we recover the 10 infinitesimal symmetries of the Poincar\'e group in a discrete theory by revising our intuition about Poincar\'e symmetry.  Our original motivation toward this effort is practical: We would like to simulate physical theories without sacrificing any of the symmetries that define such theories.  In particular, we seek to recover in discrete algorithms the conserved currents that arise from the infinitesimal Poincar\'e symmetries of continuous spacetime.  What emerges is not an algorithmic approximation of a continuous theory, but a discrete theory whose dynamics are the very algorithms for which we strive.

Toward this end, we modify the notion of spacetime, replacing the Poincar\'e-transformable spacetime vacuum with a more elementary foundation---a discrete lattice.  In contrast to previous efforts toward a lattice gauge theory of gravity, this lattice is not regarded as a structure immersed within spacetime, but rather as the framework on which spacetime is constructed.  Spacetime---as we usually think of it---no longer exists in this theory; rather, its information is encoded in fields defined on the lattice. In this sense, the lattice of our theory is better regarded as a data structure than a manifold with dimensionality and extent.  The mathematical technology for its physics will be prescribed by the \emph{discrete principal Poincar\'e bundle}.

The Poincar\'e gauge field is distinct from other Yang-Mills fields \cite{yang_conservation_1954} of the Standard Model due to its non-compact group structure and---as we shall discuss---the unique formulation of its curvature.  In this work, we nonetheless show that the `targets' of Poincar\'e transformations may be aligned with the targets of the other fundamental forces.  In particular, whereas Poincar\'e generators are most often regarded as acting on spacetime itself, we lift the Poincar\'e group action so that it transforms only the vertical,\footnote{Here, `vertical' is meant in the sense of a \emph{jet space}, whose horizontal, independent variables $X$ and vertical, dependent variables $U$ form $\ws{Jet}(0)=X\times U$. See \cite{glasser_continuous5vectortheory}.} dependent fields of our model---just as the Standard Model's ${SU(3)\times SU(2)\times U(1)}$ group acts only on vertical matter fields.

In a companion paper \cite{glasser_continuous5vectortheory}, we demonstrate that this `Poincar\'e lift' may be accomplished on a \emph{continuous} Cartesian background by replacing the familiar scalar field $\phi(x)$ with a \emph{5-vector} matter field
\begin{eqn}
\phi\left(x\right)&\rightarrow\gv{\phi}\left(x\right)\defeq\left[\begin{matrix}[l]\phi^\mu\\~\phi\end{matrix}\right]\left(x\right),
\end{eqn}
and by augmenting the $4\times4$ vierbein matrix $e_\mu^{~a}(x)$ into a \emph{solder field} we call the \emph{f\"unfbein}, a $5\times5$ matrix defined at each point of the continuous background:
\begin{eqn}
e_\mu^{~a}\left(x\right)&\rightarrow\v{e}\left(x\right)\defeq\left[\begin{matrix}e_\mu^{~a}&\v{0}\\e_\mu&1\end{matrix}\right]\left(x\right).
\end{eqn}

These more detailed fields effect a data transfer from the base space of the principal bundle to the dynamical fields in its vertical fibers.  In this paper, we show that this data transfer unburdens the base space of our theory, affording its lattice discretization without breaking Poincar\'e symmetry.  Although we replace the continuous coordinate dependence of our companion paper's fields---$f(x)$---with a discrete lattice vertex dependence---$f\n$---Poincar\'e symmetry is yet preserved.

More practically minded, discrete algorithms derived from our theory will conserve linear and angular energy-momentum (as well as the other Noether currents of the Standard Model) to machine precision.  In effect, our theory is designed for computation, rather than for pen and paper analysis.

Our approach to this problem is prompted by the observation that, in spacetime-continuum theories, \emph{independent} spacetime coordinates---acting as a Poincar\'e-set---readily admit models that conserve energy and momentum.  Such models are the workhorse of contemporary physics; even general relativity (GR) \cite{einstein_relativity_1916}, with its dynamical metric, has non-dynamical, Poincar\'e-transformable spacetime coordinates.  On the other hand, to construct a discrete, Poincar\'e-invariant theory, \emph{dependent} dynamical variables are the prerequisite targets for Poincar\'e group actions.\footnote{Such a conclusion seems inevitable after a study of the \emph{variational complex} detailed in Refs. \cite{olver_textbook_1993},\cite{hydon_variational_2004}---though we have made no attempt at a proof.}

This view---on the necessity of \emph{dependent} Poincar\'e-sets in discrete theories---was promoted in a 1986 speech of T.D. Lee's.  His speech concluded, ``I suggest that this discrete formulation might be more fundamental."\cite{lee_difference_1987}  We proceed with this suggestion in mind.

\section{Ungauged Lattice 5-Vector Theory}
Having introduced \emph{5-vector field theory} on a continuous Cartesian background in our companion paper, we now explore a discrete setting for this theory.  We begin with a discussion of the \emph{discrete principal Poincar\'e bundle}, defined with a hypercubic lattice base space.  We then develop an ungauged discrete 5-vector theory, which is counterpart to a scalar theory in flat spacetime.  In later sections, we evolve this effort into a gravitational theory by introducing a lattice Poincar\'e gauge field.

\subsection{Mathematical Preliminaries}
\subsubsection{The discrete principal bundle and its structure group}
\begin{figure}[t!]
\begin{tikzpicture}[baseline= (a).base]
\node[scale=1.3] (a) at (0,0){
\begin{tikzcd}
P \arrow[d,"\pi" ' ] \arrow[loop left]{l}{\substack{G\tleft\\\text{free}}}\\
	\left\{\v{n}\right\} \arrow[u, bend right=50,"\sigma" ' ]
\end{tikzcd}
};
\end{tikzpicture}
\caption{A depiction of the discrete principal bundle $(P,\pi,\v{n})$ with its free left $G$-action, and a choice of section $\sigma\in\Gamma(P)$.  The latter is defined such that its composition with the projection $\pi$ recovers the identity map on $\{\v{n}\}$, i.e. $\pi\circ\sigma=\ms{id}$.}
\label{Fig:PrincipalBundle}
\end{figure}
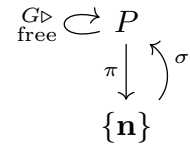

To set the stage, we define a trivial principal bundle with a discrete `base manifold', which we denote $(P,\pi,\{\v{n}\})$---see Fig. \ref{Fig:PrincipalBundle}.  The base manifold\footnote{As a colloquialism, we will continue to refer to the hypercubic lattice base space as a `base manifold', error that it is.} is taken to be a hypercubic lattice, with integer-labeled vertices $\v{n}=(n_t,n_x,n_y,n_z)$ and corresponding edges, faces, cells, and hypercells.  For simplicity, $P$ is taken to have a trivial product structure with respect to its Poincar\'e structure group---that is,
\begin{eqn}
P&=\{\v{n}\}\times G^+\\
&\defeq\{(n_t,n_x,n_y,n_z)\}\times\left\{\mR^{3,1}\rtimes SO^+(3,1)\right\}.
\end{eqn}
We denote by $G^+$ the connected subgroup of the Poincar\'e group, and we correspondingly call $P$ the \emph{discrete principal $G^+$-bundle}.  Here a point $p\in P$ is specified by $p=(\v{n},g)$, which represents a vertex assignment $\v{n}$ in the base manifold, and a choice $g\in G^+$ in the group fiber:
\begin{eqn}
g=\left(\Lambda^\mu_{~\nu},\vphi_\nu\right)=\exp\Big(\gamma_\alpha P^\alpha+\omega_{\alpha\beta}M^{\alpha\beta}\Big)^\mu_{~~\raisemath{1.8pt}{\nu}}
\end{eqn}
for some $\left\{\gamma_\alpha,\omega_{\alpha\beta}\right\}\in\mR$.  $P^\alpha$ and $M^{\alpha\beta}$ are, respectively, the translation and Lorentz transformation generators of the Poincar\'e Lie algebra.  Note that we have already separated the discrete lattice of the base manifold from the continuous Poincar\'e-sets that exist `above' the lattice, in the fibers.

We define the free left $G^+$-action $\tleft$ on an arbitrary point $p=(\v{n},g)\in P$ as
\begin{eqn}
g'\tleft p=g'\tleft(\v{n},g)\defeq\left(\v{n},g'\bullet g\right)
\end{eqn}
where $\bullet$ denotes the usual group operation of $G^+$
\begin{eqn}
g'\bullet g&=\left(\Lambda',\vphi'\right)\bullet\left(\Lambda,\vphi\right)=\left({\Lambda'}^\mu_{~\sigma}{\Lambda}^\sigma_{~\nu}~,~{\vphi'}_\sigma{\Lambda}^\sigma_{~\nu}+{\vphi}_\nu\right),
\end{eqn}
with identity element $g_e=(\mOne,\v{0})$, and where $g^{-1}\in G^+$ is given by
\begin{eqn}
g^{-1}&=\left(\Lambda,\vphi\right)^{-1}=\left(\Lambda^{-1},-\vphi\Lambda^{-1}\right).
\end{eqn}

In particular, we will employ the following faithful, non-unitary $5\times5$ matrix representation $\rho$ of the Poincar\'e group:
\begin{eqn}
\rho:\left\{
\begin{alignedat}{3}
&(\Lambda,\vphi)&&\rightarrow~~\gv{\Lambda}&&\equiv\left[\begin{matrix}\Lambda^\mu_{~\nu}&\v{0}\\\vphi_\nu&1\end{matrix}\right]\\
&(\mOne,\v{0})&&\rightarrow~~\gv{\Lambda}_0&&\equiv\left[\begin{matrix}\delta^\mu_\nu&\v{0}\\\v{0}&1\end{matrix}\right]\\
&(\Lambda,\vphi)^{-1}&&\rightarrow~~\gv{\Lambda}^{-1}&&\equiv\left[\begin{matrix}(\Lambda^{-1})^\mu_{~\nu}&\v{0}\\-\vphi_\sigma(\Lambda^{-1})^\sigma_{~\nu}&1\end{matrix}\right].
\end{alignedat}\right.
\end{eqn}

Lastly, on our trivial bundle $P$, we note that the projection map $\pi:P\rightarrow\{\v{n}\}$ is simply given by:
\begin{eqn}
\pi(p)=\pi\big((\v{n},g)\big)\defeq\v{n}.
\end{eqn}

\subsubsection{The associated fibers---the solder and matter fields---\\and their group transformations}
For the solder field of our theory, we define a $G^+$-equivariant, vector-space-valued function---(namely, a $5\times5$ matrix-valued function)---denoted $\v{e}$, which is defined at each point of the principal bundle:\footnote{The choice of the notation $F$ for the target of a group-equivariant function derives from the correspondence of such a function with the section of an associated bundle $P_F$ with fiber $F$.  $F$ is required to be vector-space-valued to enable its use in the discrete covariant derivative, which will be defined in the gauged version of our theory.}
\begin{eqn}
\v{e}:P&\rightarrow F_\v{e}\\
p&\mapsto\v{e}(p)\defeq\left[\begin{matrix}e_\mu^{~a}&\v{0}\\e_\mu&1\end{matrix}\right]\left(p\right).
\label{defineELatticeField}
\end{eqn}
Here, we define $G^+$\emph{-equivariance} in the usual way:
\begin{eqn}
\v{e}(g\tleft p)=\v{e}(p)\tright g^{-1}.
\label{gEquivariance}
\end{eqn}
We refer to the field $\v{e}$ as the \emph{f\"unfbein}, and we define\footnote{Note that we define our solder field as a zero-form---a function of $p\in P$---rather than as a one-form with arguments in the tangent bundle $TP$.  Still, we note that the lattice index $a$ appears in (the vierbein components of) the f\"unfbein, which demonstrates how it can be associated to a function over $TP$, if desired.} the right action $\tright$ in Eq.~(\ref{gEquivariance}) as follows:\footnote{At times we suppress the $p$-dependence of our fields---e.g. ${e_\mu^{~a}\equiv e_\mu^{~a}(p)}$---for compactness of notation.}
\begin{eqn}
\v{e}(p)\tright g^{-1}&=\v{e}(p)\tright\left(\Lambda,\vphi\right)^{-1}\\
&\defeq\left[\begin{matrix}e_\mu^{~a}&\v{0}\\e_\mu&1\end{matrix}\right]\cdot\left[\begin{matrix}\Lambda^\mu_{~\nu}&\v{0}\\\vphi_\nu&1\end{matrix}\right]^{-1}\\
&=\left[~\begin{matrix}[c|c]e_\mu^{~a}(\Lambda^{-1})^\mu_{~\nu}&\v{0}\\\hline e_\mu(\Lambda^{-1})^\mu_{~\nu}-\vphi_\sigma(\Lambda^{-1})^\sigma_{~\nu}&1\end{matrix}~\right].
\label{defineELatticeFieldAction}
\end{eqn}

For our matter field, we define a $G^+$-equivariant, vector-space-valued function $\gv{\phi}$ defined at each point of the principal bundle:
\begin{eqn}
\gv{\phi}:P&\rightarrow F_{\text{\tiny{$\gv{\phi}$}}}\\
p&\mapsto\gv{\phi}(p)\defeq\left[\begin{matrix}[l]\phi^\mu\\~\phi\end{matrix}\right]\left(p\right).
\end{eqn}
We define $\gv{\phi}$ to take values in the vector space of \emph{5-vectors}, which we define by their transformation under the following left action $\tleft$ of the Poincar\'e group:\footnote{Drawing an analogy between $\phi^\mu$ and the partial derivative $\partial^\mu\phi$ of scalar theory, we note that the translation group action on the \emph{scalar component} of the 5-vector is seemingly a first-order truncation of its scalar theory counterpart:
\begin{equation*}
\phi(x)\rightarrow\phi(x+\vphi)\approx\phi(x)+\vphi_\mu\partial^\mu\phi(x)+\cdots
\end{equation*}
}
\begin{eqn}
\gv{\phi}(g\tleft p)=g\tleft\gv{\phi}(p)&=(\Lambda,\vphi)\tleft\gv{\phi}(p)\\
&\defeq\left[\begin{matrix}\Lambda^\mu_{~\nu}&\v{0}\\\vphi_\nu&1\end{matrix}\right]\cdot\left[\begin{matrix}\phi^\nu\\\phi\end{matrix}\right]\\
&=\left[\begin{matrix}\Lambda^\mu_{~\nu}\phi^\nu\\\phi+\vphi_\nu\phi^\nu\end{matrix}\right].
\label{definePhiLatticeFieldAction}
\end{eqn}
We note that ``$G^+$-equivariance" is inversely defined\footnote{The inverse group actions on the matter and solder fields encode the familiar dichotomy of the `active' and `passive' notions of spacetime symmetry transformations.} for the matter field $\gv{\phi}$ relative to the more conventional equivariance of the solder field, defined in Eq.~(\ref{gEquivariance}).

We further define the 5-vector $\gv{\phi}$'s antiparticle---the \emph{twisted 5-vector}
\begin{eqn}
\gv{\tphi}:P&\rightarrow F_{\text{\tiny{$\gv{\tphi}$}}}\\
p&\mapsto\gv{\tphi}(p)\defeq\left[\begin{matrix}[l]\tphi^\mu~~\tphi\end{matrix}\right]\left(p\right)
\end{eqn}
---as well as its $G^+$-equivariant right group action:
\begin{eqn}
\gv{\tphi}(g\tleft p)=\gv{\tphi}(p)\tright g^T&=\gv{\tphi}(p)\tright(\Lambda,\vphi)^T\\
&\defeq\left[\begin{matrix}[l]\tphi^\nu~~\tphi\end{matrix}\right]\cdot\left[\begin{matrix}\Lambda^\mu_{~\nu}&\vphi_\nu\\\v{0}&1\end{matrix}\right]\\
&=\left[\tphi^\nu\Lambda^\mu_{~\nu}~\Big|~\tphi+\vphi_\nu\tphi^\nu\right].
\label{defineAPhiLatticeFieldAction}
\end{eqn}
For ease of notation, we leave the transposes of \emph{components} implicit in the above expressions---i.e., $(\tphi^\nu)^T$, $(\Lambda^\mu_{~\nu})^T$ and $(\vphi_\nu)^T$.

\subsubsection{The symmetry generators}
Because $P$ has a trivial product structure, we can globally separate the `vertical' and `horizontal' components of its tangent bundle $TP$ (and cotangent bundle $T^*P$).  We note that, like $P$, $TP=\{\v{n}+\frac{\hat{a}}{2}\}\times TG^+$ has a half-discrete, half-continuous structure.  Here, we denote by $\v{n}+\frac{\hat{a}}{2}$ the edge linking vertices $\v{n}$ and $\v{n}+\hat{a}$.  We emphasize that $\hat{a}\in\{\hat{t},\hat{x},\hat{y},\hat{z}\}$ represents not a unit tangent vector on a spacetime manifold, but a \emph{lattice direction}.  The lattice edges themselves form the basis for horizontal vector fields on $P$.  Similarly, ${\text{1-forms}}$ on the lattice are, in a sense, simpler than they are on a continuous manifold: They are single-valued along each edge and their values need not (indeed, cannot) vary smoothly from one edge to the next.

We recall that the Poincar\'e Lie algebra $\mathfrak{g}^+\cong T_{g_e}G^+$ is specified by 10 generators $\left\{P^\alpha,M^{\alpha\beta}\right\}_{\alpha\neq\beta\in\{0,1,2,3\}}$
satisfying the following Lie brackets:
\begin{eqn}
\left\llbracket P^\alpha,P^\beta\right\rrbracket&=0\\
\left\llbracket M^{\alpha\beta},P^\mu\right\rrbracket&=\eta^{\alpha\mu}P^\beta-\eta^{\beta\mu}P^\alpha\\
\left\llbracket M^{\alpha\beta},M^{\mu\nu}\right\rrbracket&=\eta^{\alpha\mu}M^{\beta\nu}+\eta^{\beta\nu}M^{\alpha\mu}-\eta^{\alpha\nu}M^{\beta\mu}-\eta^{\beta\mu}M^{\alpha\nu}
\label{PoincareLieAlgebra}
\end{eqn}
where we define $\eta^{\mu\nu}$ to have signature ($-$$+$$+$$+$).  Though our theory in this section is as yet ungauged, we note that in later sections we shall define a $\mathfrak{g}^+$-valued connection 1-form, whose pullback to the base manifold takes its values along each edge of the lattice.  We delay this discussion for now, focusing instead on global gauge transformations.

With our mathematical technology in place, we now turn to a Lagrangian description of the physics.

\subsection{Physics of the Ungauged Discrete Action}
We define the ungauged lattice 5-vector theory's discrete action on $P$ as follows:\footnote{We note that this action assumes the fields $\gv{\phi}\n$ and $\v{e}\n$ are in fact defined on the base manifold as functions of $\v{n}$---and not as functions of $p\in P$.  As usual in Yang-Mills theory, these definitions may be facilitated by the pullback of a section of $P$---${\sigma\in\Gamma(P)}$---to $\{\v{n}\}$.  While important, such mathematical rigor will not be required in this work, though the reader may consult \cite{bleecker_textbook_1981} or \cite{fredric_schuller_local_nodate}.  From hereon, we shall follow the more conventional description of gauge theories in physics, by abstracting away the principal bundle underlying the physics.  As we did for fields on $P$, we shall often use the notation $f\equiv f\n$, suppressing the lattice coordinate for the field $f$, when evaluated at $\v{n}$.}
\begin{eqn}
\ms{S}&\defeq\sum_\v{n}\ms{L}\n\\
\ms{L}\n&\defeq\gv{\tphi}\n\v{e}^T\n\gv{\ws{d}}_m\v{e}\n\gv{\phi}\n\\
&\defeq\left[\tphi^\mu~\tphi\right]\left[\begin{matrix}e_\mu^{~a}&e_\mu\\\v{0}&1\end{matrix}\right]\left[\begin{matrix}\eta_{ab}&-\md_a^+\\-\md_b^-&m^2\end{matrix}\right]\left[\begin{matrix}e_\nu^{~b}&\v{0}\\e_\nu&1\end{matrix}\right]\left[\begin{matrix}\phi^\nu\\\phi\end{matrix}\right]\\
&=\tphi^\mu g_{\mu\nu}\phi^\nu-\tphi^\mu e_\mu^{~a}\md_a^+\phi_0-\tphi_0\md_b^-(e_\nu^{~b}\phi^\nu)+m^2\tphi_0\phi_0,
\label{theAction}
\end{eqn}
where we have defined the gauge-invariant quantities:
\begin{eqn}
\phi_0\n&\defeq\phi\n+e_\mu\n\phi^\mu\n\\
\tphi_0\n&\defeq\tphi\n+e_\mu\n\tphi^\mu\n.
\label{gaugeInvQuantities}
\end{eqn}

Several notes are in order regarding this discrete action.  First, we have employed the lattice \emph{difference operators}\footnote{The lattice \emph{shift operator} $S^\v{m}$ is defined such that
\begin{equation*}
S^\v{m}f(\v{n})g(\v{n})=f(\v{n}+\v{m})g(\v{n}).
\end{equation*}
We shall always include parentheses if a lattice operator---i.e., $\{S^{\pm a},\md^\pm_a\}$---acts on more than just the object immediately to its right, except in matrix expressions---as in Eq.~(\ref{theAction}).  We further note that shift operators commute: ${S^aS^b=S^bS^a}$.}
\begin{eqn}
\md_a^\pm f\n&\defeq\pm\left(S^{\pm a}-\mOne\right)f\n=\pm\Big(f[\v{n}\pm\hat{a}]-f\n\Big),
\end{eqn}
to construct the matrix lattice operator $\gv{\ws{d}}_m$ in Eq.~(\ref{theAction}).

Second, we have, for now, forgone any notion of a volume measure---(e.g., $\md^4x\sqrt{-g}$).  We regard the Lagrangian as a real number at each vertex of the lattice base manifold; neighboring vertices are simply separated by a dimensionless integer: $1$.  In particular, $\md_a^\pm x^b=\delta^b_a$.

We use the Einstein summation convention to sum over repeated lattice indices,
\begin{eqn}
X_aY^a\defeq\smashoperator{\sum\limits_{a\in\{t,x,y,z\}}}X_aY^a=\smashoperator{\sum\limits_{a,b\in\{t,x,y,z\}}}X^aY^b\eta_{ab},
\end{eqn}
and Latin indices may be raised and lowered with the Minkowski metric $\eta_{ab}$ of signature ($-$$+$$+$$+$).  Greek indices may be raised and lowered with the metric ${g_{\mu\nu}\n\equiv\eta_{ab}e_\mu^{~a}\n e_\nu^{~b}\n}$, although it will become clear in the later gauged theory that the vierbein is best regarded as Lorentz-valued in the ungauged theory, such that ${g_{\mu\nu}\n=\eta_{\mu\nu}}$ $\forall$ $\v{n}$.

Let us derive the equations of motion (EOM) from the action in Eq.~(\ref{theAction}).  At this point we leave our theory ungauged, which leads us to make the following \emph{ungauged assumptions} for the f\"unfbein solder field:
\begin{enumerate}[label=(\alph{enumi})]
\item $\v{e}\n$ is constant at every $\v{n}$, so that its \emph{finite differences vanish} ($\md_a^\pm\v{e}\n=0$);
\item $\v{e}\n$ transforms under \emph{global} Poincar\'e gauge transformations; and
\item $\v{e}\n$ is a \emph{dynamical} variable (though its EOM will turn out to be indeterminate in the ungauged theory).
\end{enumerate}

With these assumptions for the solder field, we find the variation of our action by applying the discrete Euler operators of our matter fields, as defined by Ref.~\cite{hydon_variational_2004}:
\begin{eqn}
\ws{E}_q(\ms{L}\n)\defeq\sum\limits_\v{m}S^{-\v{m}}\pd{\ms{L}\n}{q[\v{n}+\v{m}]},
\label{eulerOperator}
\end{eqn}
where the sum over $\v{m}$ includes all lattice vertices.  We define the following gauge-invariant quantities
\begin{eqn}
(\circ)&\defeq m^2\phi_0\n-\md_b^-(e_\nu^{~b}\n\phi^\nu\n)\\
(\bullet)&\defeq m^2\tphi_0\n+\md_b^-(e_\mu^{~b}\n\tphi^\mu\n)\\
(\circ\circ)_a&\defeq\eta_{ab}e_\nu^{~b}\n\phi^\nu\n-\md_a^+\phi_0\n\\
(\bullet\bullet)_a&\defeq\eta_{ab}e_\mu^{~b}\n\tphi^\mu\n+\md_a^+\tphi_0\n,
\end{eqn}
and derive our EOM from $\ms{L}\n$, as follows:
\begin{eqn}
\hspace{-5pt}&0=\ws{E}_{\tphi^\sigma}(\ms{L}\n)=e_\sigma^{~a}\n(\circ\circ)_a+e_\sigma\n(\circ)\\
\hspace{-5pt}&0=\ws{E}_{\tphi}(\ms{L}\n)=(\circ)\\
\hspace{-5pt}&0=\ws{E}_{\phi^\sigma}(\ms{L}\n)=e_\sigma^{~a}\n(\bullet\bullet)_a+e_\sigma\n(\bullet)\\
\hspace{-5pt}&0=\ws{E}_{\phi}(\ms{L}\n)=(\bullet)\\
\hspace{-5pt}&0=\ws{E}_{e_\sigma}(\ms{L}\n)=\tphi^\sigma\n(\circ)+\phi^\sigma\n(\bullet)\\
\hspace{-5pt}&0=\ws{E}_{e_\sigma^{~a}}(\ms{L}\n)=\tphi^\sigma\n(\circ\circ)_a+\phi^\sigma\n(\bullet\bullet)_a.
\label{FiveVectorEOM}
\end{eqn}
We see that all of our EOM are solved when:
\begin{eqn}
(\circ)=(\bullet)=(\circ\circ)_a=(\bullet\bullet)_a=0.
\label{theBriefUngaugedEOM}
\end{eqn}

The EOM of ungauged 5-vector theory are analogous to the EOM of scalar field theory in flat spacetime.  We note that \emph{both} $\phi_0$ and $\tphi_0$ obey discrete Klein-Gordon equations on shell:
\begin{eqn}
0=(\md^2-m^2)\phi_0\n=(\md^2-m^2)\tphi_0\n,
\end{eqn}
where ${\md^2\equiv\eta^{ab}\md_a^+\md_b^-}$.  Despite this identical behavior, the $(\bullet)$ and $(\bullet\bullet)_a$ EOM for $\gv{\tphi}$ have an important sign difference with respect to $(\circ)$ and $(\circ\circ)_a$.  The \emph{internal} dynamics of the components of $\gv{\tphi}$ are distinctly opposite those of $\gv{\phi}$, a fact which supports $\gv{\tphi}$'s interpretation as the antiparticle of $\gv{\phi}$.

We observe that our solder field's 20 degrees of freedom (DOF) are indeterminate in the EOM of Eq.~(\ref{theBriefUngaugedEOM}).  This is to be expected in an ungauged theory---inasmuch as the `dynamics' of $\eta_{\mu\nu}$ are `indeterminate' in a flat-spacetime scalar theory.\footnote{The indeterminacy of $\v{e}$ does not hinder the simulation of our ungauged theory; after all, the solder field may simply be chosen to take a nonsingular constant value $\forall$ $\v{n}$.} The solder field's indeterminacy will be lifted, and gravity introduced, when we add gauge curvature to our Lagrangian.

Let us now describe the infinitesimal Poincar\'e symmetry of our discrete theory.  The matrix generators of the Poincar\'e group are given by:
\begin{eqn}
[P^\alpha]^\mu_{~\sigma}&\defeq\left[\begin{matrix}\mZero&\v{0}\\\delta^\alpha_\sigma&0\end{matrix}\right]\\
[M^{\alpha\beta}]^\mu_{~\sigma}&\defeq\left[\begin{matrix}\left(\delta^\alpha_\sigma\eta^{\beta\mu}-\delta^\beta_\sigma\eta^{\alpha\mu}\right)&\v{0}\\\v{0}&0\end{matrix}\right].
\label{MatrixFormalismGenerators}
\end{eqn}
We apply these generators to $\ms{L}\n$ as we would differential operators, and we observe that their prolongations transform our matrix fields wherever those fields appear---even under a difference operator $\md_a^\pm$.  In particular, each of these matrix generators---and their appropriate negations and transposes, as determined by the group actions of Eqs.~(\ref{defineELatticeFieldAction})-(\ref{defineAPhiLatticeFieldAction})---applied to the fields of the Lagrangian \emph{in situ}, generate the same transformations as the following prolonged vector fields, respectively:
\begin{eqn}
\ws{pr}[P^\alpha]&=\smashoperator{\sum\limits_{\v{n}}}\Big[\phi^\alpha\n\partial_{\phi\n}+\tphi^\alpha\n\partial_{\tphi\n}-\partial_{e_\alpha\n}\Big]\\
\ws{pr}[M^{\alpha\beta}]&=\smashoperator{\sum\limits_{\v{n}}}\Big[\phi^\sigma\n(\delta^\alpha_\sigma\eta^{\beta\nu}-\delta^\beta_\sigma\eta^{\alpha\nu})\partial_{\phi^\nu\n}\\
&\hspace{30pt}+\tphi^\sigma\n(\delta^\alpha_\sigma\eta^{\beta\nu}-\delta^\beta_\sigma\eta^{\alpha\nu})\partial_{\tphi^\nu\n}\\
&\hspace{30pt}-e_\sigma\n(\delta^\alpha_\nu\eta^{\beta\sigma}-\delta^\beta_\nu\eta^{\alpha\sigma})\partial_{e_\nu\n}\\
&\hspace{30pt}-e_\sigma^{~a}\n(\delta^\alpha_\nu\eta^{\beta\sigma}-\delta^\beta_\nu\eta^{\alpha\sigma})\partial_{e_\nu^{~a}\n}\Big].
\end{eqn}

Given the Poincar\'e invariance of the terms $\gv{\tphi}\v{e}^T$ and $\v{e}\gv{\phi}$---evident from Eqs.~(\ref{defineELatticeFieldAction}), (\ref{definePhiLatticeFieldAction}) and (\ref{defineAPhiLatticeFieldAction})---$\ms{L}\n$ is invariant under Poincar\'e transformations.  In particular:
\begin{eqn}
\ws{pr}[P^\alpha](\ms{L})&=0\\
\ws{pr}[M^{\alpha\beta}](\ms{L})&=0.
\label{variationalSymm}
\end{eqn}
Eq.~(\ref{variationalSymm}) demonstrates that our 10 infinitesimal Poincar\'e symmetries are variational symmetries of the discrete 5-vector Lagrangian.   We may therefore, in principle, pursue our theory's conservation laws via the canonical discrete Noether procedure.

The canonical Noether procedure for discrete conservation laws---detailed in Appendix A---is often more challenging to compute than its continuous counterpart; indeed, it is often easier to discover such conservation laws by inspection. \cite{hydon_p_e_difference_2014}  Nevertheless, the canonical \emph{linear} momentum Noether current can be shown to be trivial---
\begin{eqn}
A^a_{P^\alpha}=0
\label{trivialDiscreteTranslationCurrent}
\end{eqn}
---mirroring the canonical linear momentum of continuous 5-vector theory in our companion paper.\footnote{We have not pursued the discrete canonical Noether \emph{angular} momentum here, but we suspect that it, too, would be trivial.  As discussed in \cite{glasser_continuous5vectortheory}, this triviality comports with Noether's second theorem.}

Fortunately, Eq.~(67) of our companion paper's \cite{glasser_continuous5vectortheory} continuous 5-vector theory provides a natural point of departure for discovering these nontrivial currents by inspection.  We soon find:
\begin{eqn}
0=&\md_a^+\bar{T}^{a\alpha}\n\\
\defeq&\md_a^+\Bigg[e^\alpha_{~b}\eta^{bc}\bigg\{S^{-a}\Big(e_\mu^{~a}\tphi^\mu\Big)\md_c^-\phi_0\\
&\hspace{18pt}+\tphi_0S^{-a}\Big(\md_c^-(e_\mu^{~a}\phi^\mu)\Big)\bigg\}-\big(\gv{\phi}\leftrightarrow\gv{\tphi}\big)\Bigg]\\
0=&\md_a^+\bar{L}^{a\alpha\beta}\n\\
\defeq&\md_a^+\Bigg[e^\alpha_{~b}x^b[\v{n}-\hat{a}]\bar{T}^{a\beta}\n-e^\beta_{~b}x^b[\v{n}-\hat{a}]\bar{T}^{a\alpha}\n\Bigg].
\label{consLawFiveVectorMatrixTheory}
\end{eqn}

These conservation laws may be verified by the repeated application of:
\begin{enumerate}
\item our EOM as expressed in Eq.~(\ref{theBriefUngaugedEOM});
\item the discrete Leibniz rule:
\begin{eqn}
\hspace{13pt}\md_a^\pm\Big(f_1\n f_2\n\Big)=\Big(\md_a^\pm f_1\n\Big)f_2[\v{n}\pm\hat{a}]+f_1\n\md_a^\pm f_2\n;
\end{eqn}
\item the ungauged solder field assumption, which implies that ${\md_a^\pm e_\mu^{~b}\n=0}$; and
\item the relation
\begin{eqn}
\md_a^\pm\circ S^{\mp a}=\md_a^\mp.
\end{eqn}
\end{enumerate}
Validation of ${\md_a^+\bar{L}^{a\alpha\beta}\n}$ additionally requires setting ${\md_a^+\bar{T}^{a\alpha}\n=0}$ and noting the ${(\alpha\leftrightarrow\beta)}$ symmetry of ${e^\alpha_{~a}\bar{T}^{a\beta}\n}$; this symmetry is easily discovered by using the EOM to substitute for $\phi^\mu$ (or $\tphi^\mu$) in terms of $\phi_0$ (or $\tphi_0$).

Having described the Lagrangian, EOM, symmetries, and conservation laws of discrete 5-vector theory, we now introduce its Poincar\'e gauge field.

\section{Gauged Lattice 5-Vector Theory}

We now gauge our 5-vector theory to discover a lattice gauge theory of gravity.  To this end, we introduce a dynamical Poincar\'e gauge field and affine connection, which are used to define discrete covariant derivatives and gauge curvature.

We first observe that, unlike canonical gauge theories, the ungauged Lagrangian of Eq.~(\ref{theAction}) is already \emph{locally} Poincar\'e invariant.  It might appear, therefore, that our task is already done.  We quickly note, however, that because the conservation laws of Eq.~(\ref{consLawFiveVectorMatrixTheory}) require that ${\md_a^\pm e_\mu^{~b}=0}$, these conservation laws do \emph{not} hold when the solder field $\v{e}\n$ undergoes local Poincar\'e transformations.  Our motivation, then, is to gauge our preceding theory so as to restore energy-momentum conservation in a locally invariant theory.

We shall proceed in two steps: (i) We will first explore the tools of gauge theory necessary to define a gauged matter Lagrangian $\ms{L}_M\n$, and derive its 5-vector EOM---the lattice equivalent of a scalar field theory evolving in a static curved spacetime.  (ii) We will then define a pure gauge Lagrangian $\ms{L}_G\n$ and recover a discrete analog of Einstein's equations.  We thereupon discover a discrete energy-momentum conservation law of gauged 5-vector theory.

\subsection{Mathematical Preliminaries for $\ms{L}_M$}
Building upon our earlier development, we introduce the tools of gauge theory required to define a connection on our discrete principal Poincar\'e bundle $P$.

\subsubsection{The Poincar\'e gauge field}
We begin by defining two structures on our principal $G^+$-bundle that follow the conventional development of a gauge theory:
\begin{enumerate}
\item a $\mathfrak{g}^+$-valued connection one-form $\omega\big(TP,\mathfrak{g}^+\big)$; and
\item a section $\sigma:\v{n}\mapsto\left(\v{n},g_{\sigma(\v{n})}\right)$, such that $\pi\circ\sigma=\ms{id}$.
\end{enumerate}
We denote the pull-back of this connection $\mathcal{A}=\sigma^*\omega$, such that $\mathcal{A}$ is itself a $\mathfrak{g}^+$-valued one-form that is evaluated along the edges---(the `tangent bundle')---of the lattice.\footnote{Here, we adopt the discrete exterior calculus (DEC) \cite{desbrun_discrete_2005} notion of a one-form on a hypercubic lattice as taking values on its edges, or links.}  In particular, we let $\mathcal{A}_a\n$ denote the value of the connection one-form along the oriented link ${\v{n}+\frac{\hat{a}}{2}}$, and we note that ${\mathcal{A}_{-a}[\v{n}+\hat{a}]=-\mathcal{A}_a[\v{n}]}$.  We define $\mathcal{A}_a\n$ to be a dynamical $5\times5$ matrix-valued field:
\begin{eqn}
\mathcal{A}_a\n\in\ms{span}_\mR\left\{\left[P^\alpha\right],\left[M^{\alpha\beta}\right]\right\}_{\alpha\neq\beta\in\{0,1,2,3\}}
\label{matrixGeneratorNotation}
\end{eqn}
with matrices $\left[P^\alpha\right]$ and $\left[M^{\alpha\beta}\right]$ defined in Eq.~(\ref{MatrixFormalismGenerators}).  We furthermore define the $G^+$-valued \emph{comparator} matrix
\begin{eqn}
\mathcal{U}_a[\v{n}]\defeq\exp(\mathcal{A}_a[\v{n}]),
\label{signConventionForU}
\end{eqn}
along $\v{n}+\frac{\hat{a}}{2}$ and note that ${\mathcal{U}_{-a}[\v{n}+\hat{a}]=\mathcal{U}_a[\v{n}]^{-1}}$.

More concretely, we define $\Gamma^\mu_{~\nu a}\n$ as an $\mathfrak{so}(3,1)$-valued one-form and $\epsilon_{\nu a}\n$ as a $\mathfrak{t}(4)$-valued one-form, which form the Lorentz and translation components, respectively, of the $\mathfrak{g}^+$-valued gauge field $\mathcal{A}_a\n$:
\begin{eqn}
\mathcal{A}_a\n\defeq\left[\begin{matrix}\Gamma^\mu_{~\nu a}&\v{0}\\\epsilon_{\nu a}&0\end{matrix}\right]\n~\xrightarrow{\exp}~\mathcal{U}_a\n\defeq\left[\begin{matrix}\Lambda^\mu_{~\nu a}&\v{0}\\\vphi_{\nu a}&1\end{matrix}\right]\n.
\label{PoincareAlgebraAndGroupNotation}
\end{eqn}

We note that the comparator $\mathcal{U}_{\pm a}\n$ transforms as usual to enforce local Poincar\'e symmetry in our lattice gauge theory:
\begin{eqn}
\gv{\phi}\n&\rightarrow g\n\gv{\phi}\n\\
\gv{\tphi}\n&\rightarrow \gv{\tphi}\n g\n^T\\
\mathcal{U}_{\pm a}\n&\rightarrow g\n\mathcal{U}_{\pm a}\n g[\v{n}\pm\hat{a}]^{-1}
\label{comparatorTransformation}
\end{eqn}
where $\{g\n,g[\v{n}\pm\hat{a}]\}\in G^+$ are local gauge transformations at their respective lattice vertices.

\subsubsection{The affine connection}

We have already encountered aspects of our 5-vector theory that depart from the canonical development of Yang-Mills gauge theories.  In particular, our ungauged Lagrangian $\ms{L}\n$ of Eq.~(\ref{theAction}) already enjoys local Poincar\'e symmetry---a feature of gauge theories usually accomplished only after the introduction of a gauge field.  For this reason, the Poincar\'e gauge field we have introduced above merely formalizes the description of this local symmetry.

As we previously noted, however, the conserved currents in Eq.~(\ref{consLawFiveVectorMatrixTheory}) reflect only the global Poincar\'e symmetry of $\ms{L}\n$.  To render a theory with local symmetry and conservation laws simultaneously, we are compelled to introduce another gauge field beyond ${\Gamma^\mu_{~\nu a}\n}$ in our 5-vector Poincar\'e gauge theory---the affine connection ${\Gamma^b_{~ca}\n}$.

Although the affine connection is a core feature of GR, it appears in our Poincar\'e gauge theory as something of a `noncanonical' feature.  Nonetheless, we are led to develop the affine connection by the desire to relax the ${\md_a^+\v{e}=0}$ assumption of our ungauged theory while preserving energy-momentum conservation; in particular, we will jointly apply the affine and Poincar\'e gauge connections to define a covariant derivative such that ${\mD_a^+\v{e}=0}$.

We adapt the affine connection formalism of \cite{hehl_general_1976,hehl_four_1980} to our present discrete theory.  We first promote the flat lattice metric $\eta_{ab}$ to its gauged counterpart:
\begin{eqn}
g_{ab}\n\defeq\eta_{\mu\nu}e^\mu_{~a}\n e^\nu_{~b}\n.
\end{eqn}
Latin (lattice) indices may be raised and lowered with $g_{ab}$, and Greek (Poincar\'e) indices may be raised and lowered with $\eta_{\mu\nu}$.  The vertical Poincar\'e fiber itself evidently has no notion of metric curvature;\footnote{Here we see why the vierbein was best regarded as Lorentz-valued in the ungauged theory, such that $g_{\mu\nu}\n=\eta_{\mu\nu}$.} as we shall see, the curvature we will associate with the Poincar\'e fibers will be that of a gauge theory, which reflects the anholonomy of the Poincar\'e gauge connection.

Pursuing a lattice analog of the $U_4$ Riemann-Cartan spacetime described in \cite{hehl_general_1976}, we assume that the affine connection $\Gamma^b_{~ca}\n$ is metric-compatible, as follows:
\begin{eqn}
0=&\mD_a^+g_{bc}\n\\
\defeq&g_{ij}[\v{n}+\hat{a}]\exp(-\Gamma^i_{~ba}\n)\exp(-\Gamma^j_{~ca}\n)-g_{bc}\n.
\label{latticeMetricCompatibility}
\end{eqn}
Here, we have defined the covariant derivative of the lattice metric, and set it to vanish.  We observe that, while ${\Gamma^\mu_{~\nu a}\n\in\mathfrak{so}^+(3,1)}$ is a Lorentz-Lie-algebra-valued 1-form, the affine connection ${\Gamma^b_{~ca}\n\in\mathfrak{gl}(4,\mR)}$ is a 1-form on the link $\v{n}+\frac{\hat{a}}{2}$ that may assume any value in $\mR^{4\times4}$.

When written in terms of the vierbein, Eq.~(\ref{latticeMetricCompatibility}) is a quadratic matrix expression of the form
\begin{eqn}
(\ms{X}\ms{A})^{-T}\ms{\eta}(\ms{X}\ms{A})^{-1}=\ms{B}^{-T}\ms{\eta}\ms{B}^{-1},
\label{quadraticMetricCompatibility}
\end{eqn}
where ${\ms{X}\equiv\exp(\Gamma^b_{~ca}\n)}$, ${\ms{A}\equiv e_\mu^{~b}[\v{n}+\hat{a}]}$, ${\ms{B}\equiv e_\mu^{~b}\n}$, and ${\ms{\eta}\equiv\eta_{\mu\nu}}$.  It is apparent from Eq.~(\ref{quadraticMetricCompatibility}) that $\Gamma^b_{~ca}\n$ is metric-compatible if and only if $\exists$ ${\ms{Y}\in SO(3,1)}$ such that
\begin{eqn}
\ms{Y}=\ms{B}^{-1}\ms{X}\ms{A}.
\label{newGaugeRelationshipToOld}
\end{eqn}
$\ms{Y}$ is therefore naturally identified with the comparator of our Lorentz gauge field---${\ms{Y}=\exp(\Gamma^\mu_{~\nu a}\n)}$---and we accordingly require that
\begin{eqn}
\exp(\Gamma^b_{~ca}\n)\defeq e_\mu^{~b}\n\exp(\Gamma^\mu_{~\nu a}\n)e^\nu_{~c}[\v{n}+\hat{a}].
\label{NewGaugeField}
\end{eqn}
Although at first $\Gamma^b_{~ca}\n$ may have appeared to introduce new DOF into our theory, the definitional relation of Eq.~(\ref{NewGaugeField}) ensures that this is not the case.

\subsubsection{Parallel transport and the covariant derivative}

We now seek $\mD_a^\pm$---a covariant generalization of $\md_a^\pm$ to be derived from the parallel transport of our fields along the links of our base manifold.  To guide our effort, we first recall that the covariant derivative of GR varies with the tensorial rank of the objects it differentiates:
\begin{eqn}
\nabla_\sigma\phi&\defeq\partial_\sigma\phi\\
\nabla_\sigma\phi^\mu&\defeq\partial_\sigma\phi^\mu+\left\{\begin{smallmatrix}\mu\\\sigma\nu\end{smallmatrix}\right\}\phi^\nu
\label{GRFieldSpecificChritoffel}
\end{eqn}
etc., where the affine connection $\left\{\begin{smallmatrix}\mu\\\sigma\nu\end{smallmatrix}\right\}$ may be defined to enforce metric-compatibility:\footnote{We employ the notation $\left\{\begin{smallmatrix}\mu\\\sigma\nu\end{smallmatrix}\right\}$ for GR's symmetric affine connection to distinguish it from our discrete lattice theory's gauge fields.}
\begin{eqn}
0=\nabla_\sigma g_{\mu\nu}=\partial_\sigma g_{\mu\nu}-\left\{\begin{smallmatrix}\tau\\\sigma\mu\end{smallmatrix}\right\}g_{\tau\nu}-\left\{\begin{smallmatrix}\tau\\\sigma\nu\end{smallmatrix}\right\}g_{\mu\tau}.
\label{metricCompatibility}
\end{eqn}
We presently adapt $\nabla_\sigma$ for our lattice gauge theory.

To form a discrete analog of $\nabla_\sigma$ from our gauge fields, we must define the parallel transport of our matter and solder fields along lattice links; furthermore, such a parallel transport must be consistent with the group-equivariant ${(g\tleft p)}$ actions of Eqs.~(\ref{defineELatticeFieldAction})-(\ref{defineAPhiLatticeFieldAction}).

We begin by defining a \emph{covariant lattice shift operator}
\begin{eqn}
\mathring{S}^\v{m}\defeq e^{-\v{m}}S^\v{m}
\label{gaugeCovariantShift}
\end{eqn}
where the notation $e^{-\v{m}}f[\v{n}+\v{m}]$ is used to denote the parallel transport of a generic field $f$ from $\v{n}+\v{m}$ to $\v{n}$.  We correspondingly define the \emph{discrete covariant derivative}:
\begin{eqn}
\mD_a^\pm f\n\defeq\pm\big[\mathring{S}^{\pm a}-\mOne\big]f\n.
\label{covariantDerivative}
\end{eqn}

We may now reinterpret Eq.~(\ref{NewGaugeField}) as defining the vanishing covariant derivative of our vierbein:
\begin{eqn}
0&=\mD_a^+e_\mu^{~b}\n\\
&=\exp(\Gamma^b_{~ca}\n)e_\nu^{~c}[\v{n}+\hat{a}]\exp(-\Gamma^\nu_{~\mu a}\n)-e_\mu^{~b}\n.
\label{vierbeinPostulate}
\end{eqn}
Eqs.~(\ref{gaugeCovariantShift})-(\ref{vierbeinPostulate}) implicitly define the parallel transport of our vierbein.  To preserve metric-compatibility, it is apparent that the \emph{lattice} indices of our solder field must transform appropriately under parallel transport---even though we omit their transformation under a Poincar\'e gauge transformation, as in Eq.~(\ref{defineELatticeFieldAction}).  We regard this distinction between gauge transformation and parallel transport as a defining property of the solder field; $\v{e}$ is in a sense \emph{defined} to have vanishing covariant derivative.

We therefore generalize Eq.~(\ref{vierbeinPostulate}) to define \emph{solder-field-compatibility} as follows:
\begin{eqn}
0=&\mD_a^\pm\v{e}\\
\defeq&\pm\Big[U_{\pm a}\n\tleft\v{e}[\v{n}\pm\hat{a}]\tright\mathcal{U}_{\pm a}\n^{-1}-\v{e}\n\Big]
\label{solderFieldCompatibility}
\end{eqn}
where we denote by (the non-calligraphic) $U_a$ a new, lattice-indexed ${GL(5,\mR)}$-valued comparator:
\begin{eqn}
A_a\n\defeq\left[\begin{matrix}\Gamma^b_{~ca}&\v{0}\\\v{0}&0\end{matrix}\right]\n~\xrightarrow{\exp}~U_a\n\defeq\left[\begin{matrix}\Lambda^b_{~ca}&\v{0}\\\v{0}&1\end{matrix}\right]\n.
\label{NewPoincareAlgebraAndGroupNotation}
\end{eqn}
The group action symbols $\tleft$ and $\tright$ in Eq.~(\ref{solderFieldCompatibility}) serve to emphasize the proper ordering of matrix multiplications.  We have omitted the translation degrees of freedom in the definitions of $A_a$ and $U_a$ in Eq.~(\ref{NewPoincareAlgebraAndGroupNotation}), a choice that we will revisit in our later discussion.

Given the $G^+$-equivariant ${(g\tleft p)}$ actions of Eqs.~(\ref{defineELatticeFieldAction})-(\ref{defineAPhiLatticeFieldAction}), and the covariant derivative of our solder field defined in Eq.~(\ref{solderFieldCompatibility}), the self-consistent parallel transport of our matter fields is uniquely determined.  The covariant derivatives of $\gv{\phi}$ and $\gv{\tphi}$ are straightforwardly defined in agreement with these group-equivariant actions as follows:
\begin{eqn}
\mD_a^\pm\gv{\phi}\n&\defeq\pm\Big[\mathcal{U}_{\pm a}\n\tleft\gv{\phi}[\v{n}\pm\hat{a}]-\gv{\phi}\n\Big]\\
\mD_a^\pm\gv{\tphi}\n&\defeq\pm\Big[\gv{\tphi}[\v{n}\pm\hat{a}]\tright\mathcal{U}_{\pm a}^T\n-\gv{\tphi}\n\Big].
\label{defineMatterCovariantDerivs}
\end{eqn}
In Eq.~(\ref{defineMatterCovariantDerivs}), the comparators $\mathcal{U}_{\pm a}$ and $\mathcal{U}^T_{\pm a}$ effect the parallel transport $e^{\mp a}$ of our shifted matter fields.  In keeping with the field-specific notion of $\nabla_\mu$ in GR---as in Eq.~(\ref{GRFieldSpecificChritoffel})---we have defined $\mD_a^\pm$ to act on each field of 5-vector theory according to its own lattice and Poincar\'e index rank.

Having defined $\mD_a^\pm$ for the matter and solder fields of 5-vector theory, we briefly explore a few extensions of this definition.  To take the covariant derivative of \emph{components} of the above fields (e.g. $\tphi$ or $e_\mu^{~a}$), we simply treat the components as if they were Poincar\'e-transformed within their $G^+$-representation (e.g. $\gv{\tphi}$ or $\v{e}$), as in Eq.~(\ref{vierbeinPostulate}).  To take the covariant derivative of a \emph{product} of fields, we act on each field according to its own gauge transformation.  As a consequence, the covariant derivative obeys a covariant discrete Leibniz rule:\footnote{For clarity, we note that---unlike ${\md_a^\pm}$ and ${D_a^\pm}$---an index on a shift operator (${S^{\pm a}}$ or ${\mathring{S}^{\pm a}}$) or parallel transport operator (${e^{\pm a}}$) is not to be summed over unless it appears in an otherwise summed expression.}
\begin{eqn}
\mD_a^\pm(f_1f_2)=(\mD_a^\pm f_1)\mathring{S}^{\pm a}f_2+f_1\mD_a^\pm f_2.
\label{discreteCovLeibniz}
\end{eqn}
As a further consequence, invariant expressions like $\phi_0$ have trivial covariant derivatives---${\mD_a^+\phi_0=\md_a^+\phi_0}$.  Indeed, parallel transport is trivial for any expression whose indices are all contracted;\footnote{We note that the scalar quantity $\phi$ of a 5-vector transforms \emph{nontrivially} under parallel transport.  In this sense, it can be understood to have an `implicit' Poincar\'e index, as did the scalar components of the $P^\alpha$ generator in Eq.~(\ref{MatrixFormalismGenerators}).  On the other hand, $\phi_0$ is completely invariant.} parallel transport is likewise \emph{nontrivial} for any expression with an uncontracted (lattice or Poincar\'e) index.

We summarize in Table \ref{tblCovShifts} the covariant shifts that arise from our definitions of parallel transport throughout this section.

\begin{table}[h!]
\normalsize
\begin{tabular}{l|l}
$Q\n$ & $\mathring{S}^{\pm a}Q\n\defeq e^{\mp a}S^{\pm a}Q\n$\\\hline\hline
 $\gv{\phi}\n$ & $\mathcal{U}_{\pm a}\n\gv{\phi}[\v{n}\pm \hat{a}]$\\
$\gv{\tphi}\n$ & $\gv{\tphi}[\v{n}\pm \hat{a}]\mathcal{U}_{\pm a}\n^T$\\
\hline
$\v{e}\n$ & $U_{\pm a}\n\v{e}[\v{n}\pm \hat{a}]\mathcal{U}_{\pm a}\n^{-1}=\v{e}\n$\\
\hline
$\mathcal{U}_b\n$ & $\mathcal{U}_{\pm a}\n\mathcal{U}_b[\v{n}\pm\hat{a}] \mathcal{U}_{\pm a}[\v{n}+\hat{b}]^{-1}$\\
$\mathcal{U}_{-b}\n$ & $\mathcal{U}_{\pm a}\n\mathcal{U}_{-b}[\v{n}\pm\hat{a}] \mathcal{U}_{\pm a}[\v{n}-\hat{b}]^{-1}$\\
\hline
$U_b\n$ & $U_{\pm a}\n U_b[\v{n}\pm\hat{a}] U_{\pm a}[\v{n}+\hat{b}]^{-1}$\\
$U_{-b}\n$ & $U_{\pm a}\n U_{-b}[\v{n}\pm\hat{a}] U_{\pm a}[\v{n}-\hat{b}]^{-1}$
\end{tabular}
\caption{For convenience in deriving our EOM, we summarize the covariant shifts that have been defined (or implied) throughout the preceding section.}
\label{tblCovShifts}
\end{table}

\subsection{Physics of the Gauged Matter Action}

We are at last prepared to gauge the matter Lagrangian of Eq.~(\ref{theAction}).  We define $\ms{\widehat{L}}_M\n$ as follows:\footnote{We employ the notation ${\ms{\widehat{L}}_M}$ to distinguish this matter Lagrangian from $\ms{L}_M$ in Eq.~(\ref{volumedL}), which includes a volumetric measure factor.}
\begin{eqn}
\ms{\widehat{L}}_M\n\defeq&\gv{\tphi}\n\v{e}^T\n\gv{\ws{D}}_m\v{e}\n\gv{\phi}\n\\
\defeq&\left[\tphi^\mu~\tphi\right]\left[\begin{matrix}e_\mu^{~a}&e_\mu\\\v{0}&1\end{matrix}\right]\left[\begin{matrix}g_{ab}&-\mD_a^+\\-\mD_b^-&m^2\end{matrix}\right]\left[\begin{matrix}e_\nu^{~b}&\v{0}\\e_\nu&1\end{matrix}\right]\left[\begin{matrix}\phi^\nu\\\phi\end{matrix}\right]\\
=&\tphi^\mu\eta_{\mu\nu}\phi^\nu-\tphi^\mu e_\mu^{~a}\mD_a^+\phi_0-\tphi_0\mD_b^-(e_\nu^{~b}\phi^\nu)+m^2\tphi_0\phi_0.
\label{theMatterAction}
\end{eqn}
By simply promoting ${\eta_{ab}}$ to ${g_{ab}}$, and ${\md_a^\pm}$ to ${\mD_a^\pm}$, we have adopted the principle of \emph{minimal coupling} to form the matter Lagrangian.  In the final equality above, we have noted that ${g_{ab}e_\mu^{~a}e_\nu^{~b}=\eta_{\mu\nu}}$.

Let us promptly derive the matter field EOM of our gauged theory, employing the gauge covariant Euler operators ${\mathring{\ws{E}}_q(\ms{\widehat{L}}_M\n)}$ for ${q\in\{\tphi^\sigma,\tphi,\phi^\sigma,\phi\}}$, as defined in Appendix B.  We discover that
\begin{eqn}
\hspace{-5pt}&0=\mathring{\ws{E}}_{\tphi^\sigma}(\ms{\widehat{L}}_M\n)=(\circledast\circledast)_\sigma+e_\sigma\n(\circledast)\\
\hspace{-5pt}&0=\mathring{\ws{E}}_{\tphi}(\ms{\widehat{L}}_M\n)=(\circledast)\\
\hspace{-5pt}&0=\mathring{\ws{E}}_{\phi^\sigma}(\ms{\widehat{L}}_M\n)=(\invcircledast\invcircledast)_\sigma+e_\sigma\n(\invcircledast)\\
\hspace{-5pt}&0=\mathring{\ws{E}}_{\phi}(\ms{\widehat{L}}_M\n)=(\invcircledast)
\label{GaugedFiveVectorEOM}
\end{eqn}
where we have symbolized the following expressions:
\begin{eqn}
(\circledast)&\defeq m^2\phi_0-\mD_b^-(e_\mu^{~b}\phi^\mu)\\
(\invcircledast)&\defeq m^2\tphi_0+\mD_b^-(e_\mu^{~b}\tphi^\mu)\\
(\circledast\circledast)_\sigma&\defeq\eta_{\sigma\mu}\phi^\mu-e_\sigma^{~a}\mD_a^+\phi_0\\
(\invcircledast\invcircledast)_\sigma&\defeq\eta_{\sigma\mu}\tphi^\mu+e_\sigma^{~a}\mD_a^+\tphi_0.
\end{eqn}
Our gauged EOM are therefore satisfied when:
\begin{eqn}
(\circledast)=(\invcircledast)=(\circledast\circledast)_\sigma=(\invcircledast\invcircledast)_\sigma=0.
\label{theBriefGaugedEOM}
\end{eqn}

We immediately note that both $\phi_0$ and $\tphi_0$ obey a discrete analog of the curved-spacetime Klein-Gordon equation:
\begin{eqn}
0=\Big[\mD_b^-g^{ba}\md_a^+-m^2\Big]\phi_0=\Big[\mD_b^-g^{ba}\md_a^+-m^2\Big]\tphi_0.
\label{5VectorCurvedSpaceMatterEOM}
\end{eqn}
We have reduced $\mD_a^+\phi_0$ to its equivalent $\md_a^+\phi_0$ in the above expression (as we have for $\tphi_0$).  However, $\mD_b^-$ cannot be reduced in the same way, given its operation on ${g^{ba}\md_a^+\phi_0}$, an expression with an uncontracted lattice index $b$ that parallel transports nontrivially via the affine connection $\Gamma^b_{~ca}$.

In the continuous limit, our matter fields thus recover the curved-spacetime dynamics of a scalar field in GR:
\begin{eqn}
0=\Big(\nabla^\mu\partial_\mu-m^2\Big)\phi.
\label{ScalarCurvedSpaceMatterEOM}
\end{eqn}

Having successfully derived the gauged EOM for the matter fields of lattice 5-vector theory, we now pursue the dynamics of its gauge and solder fields.

\subsection{Mathematical Preliminaries for $\ms{L}_G$}

\subsubsection{The gauge field curvature}

We begin by exploring the curvature of the Poincar\'e gauge field.  We note that there are no known fundamental forces associated with the translation components of the Poincar\'e group---which might, for example, give rise to \emph{longitudinal gravity waves}.  Although in principle our theory could accommodate such a force, we shall follow the example of GR and omit these degrees of freedom from our definition of curvature---as we did from our definition of $A_a\n$ in Eq.~(\ref{NewPoincareAlgebraAndGroupNotation}).

We therefore select our gauge curvature to capture only the Lorentz gauge field's anholonomy.  We find that the formalism of Einstein-Cartan (EC) gravity \cite{cartan_sur_1923,cartan_sur_1925,kibble_lorentz_1961,sciama_physical_1964} naturally accommodates the fields of our theory.  In EC gravity, the vierbein---rather the metric---plays a central role.  The affine connection in EC gravity is also free to admit torsion---${\Gamma^b_{~ca}\n\neq\Gamma^b_{~ac}\n}$---a feature that comports with the distinct roles of the affine connection's lower indices in Eq.~(\ref{NewGaugeField}).

For the purposes of our lattice theory, we therefore endeavor to define a Wilson loop \cite{wilson_confinement_1974} for the Lorentz components of the gauge field that recovers the EC action in the continuous limit.  We follow several previous efforts \cite{smolin_quantum_1979,mannion_general_1981,kondo_euclidean_1984,menotti_reflection_1986,menotti_poincare_1987,hamber_gravitational_2010,catterall_sitter_2012} that similarly define a gravitational lattice curvature, although we modify the foregoing gauge theories to suit the `standard' (or `fundamental') Lorentz representation of our gauge field.  Following these references, we employ the Ne'eman-Regge-Trautman (NRT) formalism for the EC action \cite{neeman_gravity_1978,trautman_einstein-cartan_1972}.

In the continuous limit, we would like the discrete gauge Lagrangian $\ms{L}_G$ to recover the following NRT Lagrangian:
\begin{eqn}
L\defeq&\frac{1}{8}\epsilon^{abcd}\epsilon_{\mu\nu\alpha\beta}e^\alpha_{~c}e^\beta_{~d}R^\mu_{~\sigma ab}\eta^{\sigma\nu}\\
=&\frac{e}{2}e_\mu^{~a}e_\nu^{~b}R^\mu_{~\sigma ab}\eta^{\sigma\nu}
\label{continuousCurvature}
\end{eqn}
where the second equality follows from the identity \cite{smolin_quantum_1979}
\begin{eqn}
\epsilon^{abcd}\epsilon_{\mu\nu\alpha\beta}e^\alpha_{~c}e^\beta_{~d}=2e(e_\mu^{~a}e_\nu^{~b}-e_\nu^{~a}e_\mu^{~b})
\label{smolinIdentity}
\end{eqn}
and the antisymmetry of the Riemann tensor, defined by
\begin{eqn}
R^\mu_{~\sigma ab}(x)\defeq\partial_{\raisemath{-2.2pt}{[a}}\Gamma^\mu_{~\sigma|b]}(x)+
\Gamma^\mu_{~\lambda[a}(x)\Gamma^\lambda_{~\sigma|b]}(x).
\label{continuousRiemannTensor}
\end{eqn}
In the above, $\epsilon$ denotes the Levi-Civita symbol, and\footnote{Although we will continue to refer to $e^\mu_{~a}$ as the \emph{inverse} vierbein, it is perhaps a more natural geometric object than $e_\mu^{~a}$.}
\begin{eqn}
e\defeq\ms{det}\left[e^\mu_{~a}\right]=\sqrt{-\ms{det}[g_{ab}]}\eqdef\sqrt{-g}.
\end{eqn}
In this formulation of EC gravity, the solder field and the Poincar\'e gauge field are assumed to be independent.  We furthermore observe that the translation degrees of freedom of the gauge field are omitted from this definition, as desired.

To form a discrete analog of Eq.~(\ref{continuousRiemannTensor}), we define the \emph{discrete Riemann tensor} by a Wilson loop:
\begin{eqn}
\square^\mu_{~\nu ab}\n\defeq&\Lambda^\mu_{~\lambda a}\n\Lambda^\lambda_{~\sigma b}[\v{n}+\hat{a}]\Lambda^\sigma_{~\tau a}[\v{n}+\hat{b}]^{-1}\Lambda^\tau_{~\nu b}\n^{-1}
\label{wilsonPlaquetteRiemann}
\end{eqn}
where we use the notation ${\Lambda^\mu_{~\nu a}\n\equiv\exp(\Gamma^\mu_{~\nu a}\n)}$ of Eq.~(\ref{PoincareAlgebraAndGroupNotation}), and we recall that ${\Lambda^\mu_{~\nu a}\n^{-1}=\exp(-\Gamma^\mu_{~\nu a}\n)}$.  ${\Lambda^\mu_{~\nu a}\n}$ is appropriately viewed as a \emph{Lorentz comparator}---a `subfield' of $\mathcal{U}_a\n$.  We furthermore note that, if desired, Eq.~(\ref{NewGaugeField}) may be substituted into Eq.~(\ref{wilsonPlaquetteRiemann}) to express $\square^\mu_{~\nu ab}\n$ in terms of the affine connection ${\Gamma^b_{~ca}\n}$.

Following Eq.~(\ref{continuousCurvature}), therefore, we define a discrete analog of the Einstein-Cartan Lagrangian, as follows:
\begin{eqn}
\ms{L}_G\n\defeq&\frac{1}{8}\epsilon^{abcd}\epsilon_{\mu\nu\alpha\beta}e^\alpha_{~c}\n e^\beta_{~d}\n\square^\mu_{~\sigma ab}\n\eta^{\sigma\nu}\\
=&\frac{e\n}{2}e_\mu^{~a}\n e_\nu^{~b}\n\square^\mu_{~\sigma ab}\n\eta^{\sigma\nu}\\
\eqdef\hspace{-3pt}&\hspace{3pt}\frac{e\n}{2}R\n.
\label{discreteCurvature}
\end{eqn}
As above, ${e\n\defeq\ms{det}\left(e^\mu_{~a}\n\right)}$, and we have defined ${R\n}$, a discrete analog of the Ricci scalar curvature.  The second equality of Eq.~(\ref{discreteCurvature}) follows after substituting Eq.~(\ref{smolinIdentity}), as well as the identity ${\square^\mu_{~\nu ab}\eta^{\nu\sigma}=\square^\sigma_{~\nu ba}\eta^{\nu\mu}}$.

As discussed in \cite{mannion_general_1981,menotti_reflection_1986}, we have not defined a lattice Lorentz curvature by the traditional method of a Yang-Mills theory---$\ms{Tr}\big(\Lambda_a\n \Lambda_b\left[\v{n}+\hat{a}\right]\Lambda_a[\v{n}+\hat{b}]^{-1}\Lambda_b\n^{-1}\big)$
---because such a trace over group indices would reduce to $R^\mu_{~\nu ab}R_\mu^{~\nu ab}$ in the continuous limit.  In this sense, Eq.~(\ref{discreteCurvature}) does not define a canonical gauge theory, since it requires the vierbein in its construction.

Before proceeding, we verify that Eq.~(\ref{discreteCurvature}) recovers Eq.~(\ref{continuousCurvature}) in the continuous limit.  We follow \cite{kogut_introduction_1979} Eq.~(8.7), applying the BCH formula to ${\square^\mu_{~\sigma ab}}$ to find:
\begin{eqn}
\square^\mu_{~\sigma ab}&\approx\exp\Big(\partial_a\Gamma_b-\partial_b\Gamma_a+[\Gamma_a,\Gamma_b]\Big)^\mu_{~~\raisemath{1.8pt}{\sigma}}\\
&\approx\delta^\mu_\sigma+\Big(\partial_a\Gamma_b-\partial_b\Gamma_a+[\Gamma_a,\Gamma_b]\Big)^\mu_{~~\raisemath{1.8pt}{\sigma}}+\cdots
\end{eqn}
From the first equality of Eq.~(\ref{discreteCurvature}), it is evident that the 0$^\text{th}$-order contribution to $R\n$ vanishes in the continuous limit, due to the contraction of symmetric and antisymmetric indices in ${\delta^\mu_\sigma\eta^{\sigma\nu}\epsilon_{\mu\nu\alpha\beta}}$.  The 1$^\text{st}$-order contribution recovers Eq.~(\ref{continuousCurvature}), as desired.

\subsection{Physics of the Gauged Total Action}
We are at last prepared to fully define gauged 5-vector theory.  We define our total Lagrangian to be
\begin{eqn}
\ms{L}\n\defeq\kappa\ms{L}_M\n+\ms{L}_G\n
\label{theTotalAction}
\end{eqn}
where
\begin{eqn}
\ms{L}_M\n\defeq e\n\ms{\widehat{L}}_M\n\\
\ms{L}_G\n\defeq e\n\ms{\widehat{L}}_G\n
\label{volumedL}
\end{eqn}
with ${\ms{\widehat{L}}_M\n}$ defined in Eq.~(\ref{theMatterAction}) and $\ms{L}_G\n$ defined in Eq.~(\ref{discreteCurvature}).  We have introduced a constant of proportionality $\kappa$ between our matter and gauge Lagrangians.

We now proceed to derive our gauge and solder field EOM.  This effort is assisted by a few preliminary observations.  First, we note an identity:
\begin{eqn}
\partial_{(e_\sigma^{~a})}e=-ee^\sigma_{~a},
\end{eqn}
a relation that may be derived from the well known expression ${\delta g=-gg_{ab}\delta g^{ab}}$.

Second, we note that a measure of care is required to derive EOM for our gauge field while preserving its group-theoretic properties.  If we derived the EOM for an arbitrary component of $\Gamma^\mu_{~\nu a}$, for example, we would be forced to exogenously constrain the field's evolution to the Lorentz Lie algebra.  To avoid this complication, we may solve for the EOM of the coefficients of the gauge field's generators, as described in \cite{ipp_implicit_2018}.  In particular, we may decompose our gauge field into its Lorentz generators:
\begin{eqn}
\Lambda^\mu_{~\nu a}\n&=\exp(\Gamma^\mu_{~\nu a}\n)=\exp(\omega_{\sigma\tau a}\n[M^{\sigma\tau}]^\mu_{~\nu})
\end{eqn}
where we have used brackets to denote the matrix Lorentz generators, as in Eq.~(\ref{MatrixFormalismGenerators}).  We then derive our EOM with respect to the coefficients $\omega_{\mu\nu a}\n$.  For example, we note that:
\begin{eqn}
\pd{\Lambda^\mu_{~\nu a}\n}{\omega_{\sigma\tau b}\n}=
\begin{cases}
[M^{\sigma\tau}]^\mu_{~\lambda}\Lambda^\lambda_{~\nu a}\n&~~\text{if $a=b$}\\
0&~~\text{otherwise.}
\end{cases}
\end{eqn}

Lastly, we consider how to apply the gauge covariant Euler operator $\mathring{\ws{E}}_q$ to the gauge field comparators in ${\ms{L}_G\n}$.  As expressed in Eq.~(\ref{wilsonPlaquetteRiemann}), the comparators of the Wilson loop can be regarded as `based at' $\v{n}$ and its nearest neighbors---$\{\v{n},\v{n}+\hat{a},\v{n}+\hat{b}\}$---and not at $\v{n}+\hat{a}+\hat{b}$.  When calculating $\mathring{\ws{E}}_{\omega_{\sigma\tau c}}(\ms{L}_G\n)$, therefore, we will parallel transport each comparator of the Wilson loop to $\v{n}$ by a zero- or single-link path only.  See Table \ref{tblCovShifts} for explicit definitions of our comparators' parallel transports.  In simple cases, these transports produce appealingly intuitive results.  For example:
\begin{eqn}
e^{a}\mathcal{U}_a[\v{n}]&=\mathcal{U}_a[\v{n}+\hat{a}]\\
\mathring{S}^{-a}\square^\mu_{~\sigma ab}\n&=\square^\mu_{~\sigma b,-a}\n\\
\mathring{S}^{-b}\square^\mu_{~\sigma ab}\n&=\square^\mu_{~\sigma,-ba}\n.
\label{simpleCovariantShiftExamples}
\end{eqn}

We are now in position to derive our gauge and solder field EOM:
\begin{eqn}
\hspace{-5pt}&0=\mathring{\ws{E}}_{e_\sigma}(\ms{L}\n)=e\left\{\tphi^\sigma(\circledast)+\phi^\sigma(\invcircledast)\right\}\\
\hspace{-5pt}&0=\mathring{\ws{E}}_{e_\sigma^{~a}}(\ms{L}\n)=-ee^\sigma_{~a}\left\{\kappa\ms{\widehat{L}}_M+\ms{\widehat{L}}_G\right\}+ee_\mu^{~b}\square^\mu_{~\nu ba}\eta^{\nu\sigma}\\
\hspace{-5pt}&\hspace{73pt}-\kappa e\left\{\tphi^\sigma\mD_a^+\phi_0+\tphi_0\mD_a^-\phi^\sigma\right\}\\
\hspace{-5pt}&0=\mathring{\ws{E}}_{\omega_{\alpha\beta c}}(\ms{L}\n)=\frac{1}{2}e\eta^{\nu\sigma}\\
&\hspace{50pt}\cdot\bigg\{[M^{\alpha\beta}]^\mu_{~\tau}\Big(e_\mu^{~\slashed{c}}e_\nu^{~b}\square^\tau_{~\sigma\slashed{c}b}+e_\mu^{~a}e_\nu^{~\slashed{c}}\square^\tau_{~\sigma\slashed{c},-a}\Big)\\
&\hspace{60pt}-\Big(e_\mu^{~a}e_\nu^{~\slashed{c}}\square^\mu_{~\tau a\slashed{c}}+e_\mu^{~\slashed{c}}e_\nu^{~b}\square^\mu_{~\tau,-b\slashed{c}}\Big)[M^{\alpha\beta}]^\tau_{~\sigma}\bigg\}.
\label{SolderGaugeEOM}
\end{eqn}
In the last expression, we denote the index $c$ with a slash to indicate that it is not summed over.

These EOM may be considerably simplified.  Noting that
\begin{eqn}
\ms{\widehat{L}}_M&=\tphi^\mu(\circledast\circledast)_\mu+\tphi_0(\circledast)
\end{eqn}
vanishes on shell, we may contract ${\eta_{\sigma\mu}e^\mu_{~a}}$ with ${\mathring{\ws{E}}_{e_\sigma^{~b}}(\ms{L}\n)}$ in Eq.~(\ref{SolderGaugeEOM}) to discover a discrete analog of Einstein's equations:
\begin{eqn}
&\left[e^\nu_{~a}e_\mu^{~c}\square^\mu_{~\nu cb}-\frac{1}{2}g_{ab}R\right]=\kappa e^\mu_{~a}\eta_{\mu\sigma}\Big[\tphi^\sigma\mD_b^+\phi_0+\tphi_0\mD_b^-\phi^\sigma\Big].
\label{EinsteinsEquations}
\end{eqn}
Eq.~(\ref{EinsteinsEquations}) can be reexpressed in the familiar notation of GR:\footnote{We use the notation $\mathring{T}$ to distinguish the gauged energy-momentum from its ungauged counterpart $T$---as appears in Eq.~(\ref{consLawFiveVectorMatrixTheory}).  $\bar{T}$ and $\mathring{\bar{T}}$ denote the energy-momenta symmetrized over $(\gv{\phi}\leftrightarrow\gv{\tphi})$.}
\begin{eqn}
G_{ab}\n=\kappa\mathring{T}_{ab}\n
\label{einsteinEqnSymbolic}
\end{eqn}
where
\begin{eqn}
R_{ab}\n\defeq&e^\nu_{~a}\n e_\mu^{~c}\n\square^\mu_{~\nu cb}\n\\
G_{ab}\n\defeq&R_{ab}\n-\frac{1}{2}g_{ab}\n R\n\\
\mathring{T}_{ab}\n\defeq&e^\mu_{~a}\n\eta_{\mu\sigma}\Big[\tphi^\sigma\n\mD_b^+\phi_0\n+\tphi_0\n\mD_b^-\phi^\sigma\n\Big].
\label{EinsteinTensorNotations}
\end{eqn}
We note that by allowing the exchange of lattice and Lorentz indices,
\begin{eqn}
R_{ab}\n=\square^c_{~acb}\n.
\end{eqn}

To simplify the EOM of ${\mathring{\ws{E}}_{\omega_{\alpha\beta c}}(\ms{L}\n)}$ in Eq.~(\ref{SolderGaugeEOM}), we substitute Eq.~(\ref{MatrixFormalismGenerators}) for $[M^{\alpha\beta}]^\mu_{~\tau}$ and again use the vierbein to freely exchange and rearrange indices, yielding:
\begin{eqn}
0&=g^{\raisemath{1pt}{\slashed{c}[b}}\square^{d]a}_{~~~\slashed{c}a}+g^{\raisemath{1pt}{a[b}}\square^{d]\slashed{c}}_{~~~\slashed{c},-a}\\
&=g^{\raisemath{1pt}{\slashed{c}[b}}\square^{d]a}_{~~~\slashed{c}a}+\mathring{S}^{-a}\left(g^{\raisemath{1pt}{a[b}}\square^{d]\slashed{c}}_{~~~a\slashed{c}}\right).
\end{eqn}

Lastly, we observe that although the Lorentz DOF of the solder and gauge fields now evolve nontrivially, the EOM for $e_\sigma$ in Eq.~(\ref{SolderGaugeEOM}) remains indeterminate.  This is to be expected, given our omission of the translation DOF from the definition of gauge curvature.

Having derived the gauged EOM, we now briefly examine the conservation laws of our `noncanonical' gauge theory.  Although a full exposition of our gauged conservation laws is deferred to future work, we derive a few immediate results.

We first note that the ungauged theory's conserved linear momentum $\bar{T}^{a\alpha}$ in Eq.~(\ref{consLawFiveVectorMatrixTheory}) was `symmetrized' by subtracting ${(\gv{\phi}\leftrightarrow\gv{\tphi})}$.  This made it possible to construct a conserved angular momentum ${\bar{L}^{a\alpha\beta}}$.  However, it is easily verified that an `unsymmetrized' linear momentum ${T^{a\alpha}}$ is also conserved in the ungauged theory:
\begin{eqn}
0=&\md_a^+T^{a\alpha}\n\\
\defeq&\md_a^+\Bigg[e^\alpha_{~b}\eta^{bc}\bigg\{S^{-a}\Big(e_\mu^{~a}\tphi^\mu\Big)\md_c^-\phi_0+\tphi_0S^{-a}\Big(\md_c^-(e_\mu^{~a}\phi^\mu)\Big)\bigg\}\Bigg].
\label{unsymmMomentumTensor}
\end{eqn}
$T^{a\alpha}$ of Eq.~(\ref{unsymmMomentumTensor}) is essentially the ungauged counterpart of $\mathring{T}_{ab}$ in Eq.~(\ref{EinsteinTensorNotations}), although lattice-specific aspects of their expressions distinguish them.  Nonetheless, in the continuous, ungauged (${g\rightarrow\eta}$, ${\Gamma\rightarrow0}$) limit:
\begin{eqn}
\eta^{\alpha\beta}e_\beta^{~b}g^{ca}\mathring{T}_{ab}\approx T^{c\alpha}.
\end{eqn}
In this sense, $\mathring{T}_{ab}$ appearing on the right hand side of Eq.~(\ref{einsteinEqnSymbolic}) behaves like the familiar energy-momentum tensor of GR.  On the other hand, ${R_{[ab]}\neq0}$ leads to an indicial asymmetry in $\mathring{T}_{ab}$.  This asymmetry in the energy-momentum tensor is a typical feature of EC gravity.

Furthermore, we should expect that, as in previous expositions of continuous EC gravity \cite{hehl_general_1976,hehl_four_1980,trautman_einstein-cartan_2006}, the covariant divergence of $\mathring{T}_{ab}$ will be nonvanishing in our discrete gauged theory.  Rather than attempting to define a conservation law from $\mathring{T}_{ab}$ as defined in Eq.~(\ref{EinsteinTensorNotations}), we instead `gauge' the symmetrized tensor $\bar{T}^{a\alpha}$ from Eq.~(\ref{consLawFiveVectorMatrixTheory}) of the ungauged theory.  Substituting from the EOM of Eq.~(\ref{theBriefGaugedEOM}), we calculate as follows:
\begin{eqn}
\mD_a^+\mathring{\bar{T}}^{a\alpha}\n\defeq&\mD_a^+\bigg[e^\alpha_{~b}g^{bc}\Big\{\mathring{S}^{-a}\Big(e_\mu^{~a}\tphi^\mu\Big)\mD_c^-\phi_0\\
&\hspace{18pt}+\tphi_0\mathring{S}^{-a}\Big(\mD_c^-(e_\mu^{~a}\phi^\mu)\Big)\Big\}-\big(\gv{\phi}\leftrightarrow\gv{\tphi}\big)\bigg]\\
=&e_\mu^{~a}e^\alpha_{~b}g^{bc}\bigg[\tphi_0\left[\mD_a^-,\mD_c^-\right]\phi^\mu+\tphi^\mu\left[\mD_a^+,\mD_c^-\right]\phi_0\bigg]\\
&\hspace{52pt}-\big(\gv{\phi}\leftrightarrow\gv{\tphi}\big).
\label{consLawGaugedFiveVectorMatrixTheory1}
\end{eqn}
In the continuous limit, this conservation law takes the form
\begin{eqn}
\hspace{-10pt}\mD_a^+\mathring{\bar{T}}^{a\alpha}\n&\approx e_\mu^{~a}e^\alpha_{~b}g^{bc}\Big[\tphi_0R^\mu_{~\nu ac}\phi^\nu+\tphi^\mu S^b_{~ac}\partial_b\phi_0\Big]-\big(\gv{\phi}\leftrightarrow\gv{\tphi}\big)
\label{continuousLimitGaugedConsLaw}
\end{eqn}
where \emph{Cartan's torsion tensor} $S^b_{~ac}$ is defined by the antisymmetric part of the affine connection:
\begin{eqn}
S^b_{~ac}\defeq\Gamma^b_{~ac}-\Gamma^b_{~ca}.
\end{eqn}
Eq.~(\ref{continuousLimitGaugedConsLaw}) recovers a result similar to Eq.~(4.39) of \cite{hehl_general_1976}.  In EC gravity, the gauged energy-momentum tensor is conserved up to second-order derivatives---a remainder that, due to its `non-locality', is nonetheless consistent with the equivalence principle of GR. \cite{hehl_four_1980}

We lastly observe that the conservation law in Eq.~(\ref{consLawGaugedFiveVectorMatrixTheory1}) admits \emph{local} Poincar\'e transformations, as we originally hoped for our gauged 5-vector theory.

\section{Conclusions and Questions\label{concludeSection}}

Having developed lattice 5-vector theory and demonstrated its apparent compatibility with the formalism of Einstein-Cartan gravity, we now reflect upon the physical and mathematical implications of this theory.  We allow ourselves considerable latitude in this discussion, and offer more speculative thoughts in the hope of spurring further extensions of this work in the community.

\subsection{Physical Considerations}

The foregoing Poincar\'e lattice gauge theory appears to have significant implications for our interpretation of the physical universe.  In our judgment, the most consequential suggestion of this work is the possibility that the laws of physics can be faithfully rendered on a discrete lattice.  Our redefinition of Poincar\'e symmetry is central to this evolution of our physical intuition.

We also observe the modified notions of time and space appearing in our theory.  As discussed in our companion paper, 5-vector theory apparently draws upon both conceptions of time asserted by Newton and Einstein.  In particular, our discrete lattice theory describes the coexistence of a `digital' absolute universal Newtonian clock $\{n_t\}$, and an `analog' relativistic local Einsteinian clock ${e^t_{~a}\n}$.

We may view our theory as a natural consequence of an axiom and an experimental fact:
\begin{itemize}
\item \underline{Axiom}: The universe is discrete.
\item \vspace{-6pt}\underline{Fact}: Its laws have Poincar\'e symmetry.
\end{itemize}
The discrete principal $G^+$-bundle we have constructed is an instinctive mathematical model for this axiomatic universe.  However, we ascribe no uniqueness to the discrete principal bundle in this regard.

It is worth dwelling in some detail on the nature of the discrete lattice we have defined.  Its `dimensionality' is only determined by the number of edges linking a vertex to its neighbors.  Furthermore, there is nothing \emph{within} the hypercubes of the lattice---it is not a structure embedded in some larger continuous manifold; the lattice in our theory represents the entirety of the universe.  In this spare conception, the lattice is better regarded as a data structure that contains a countable data set.  It can be deformed, and its information can be stored in any manner desired, so long as its oriented graph and the data of its fields are preserved.  In this sense, our lattice theory shares the deformability characteristic of the holographic principle \cite{t_hooft_dimensional_1993,susskind_world_1995}.

A description of the universe as a lattice---more precisely, as a discrete principal bundle---raises more questions than it answers.  Taking the notion of a discrete universe seriously, one is compelled to ask: How do the fundamental forces regulate the sizes of lattice `pixels'---that is, the values of $e^\mu_{~a}\n$?  How many pixels comprise the various structures of an atom?  Can the oriented graph of the lattice evolve?  Are there viable lattice geometries beyond the hypercubic lattice, and how would their predictions differ?  How can we observe the effects of the lattice, and at what scales would these effects be visible?

This last question raises the possibility that lattice effects might be observed on large scales in astronomical data.  For example, a pixelated universe would presumably vary in its `resolution' $\mathcal{R}\n$, which could be estimated via a local average of the volumetric measure $e\n$:
\begin{eqn}
\mathcal{R}_N\n\defeq\left[\big\langle e[\v{m}]\big\rangle_{|\v{m}-\v{n}|\leq N}\right]^{-1}.
\end{eqn}
In `low-resolution' sectors of a discrete universe, a continuous gravitational theory such as GR would predict different outcomes from experiment.  For example, a discrete theory would modify predictions for the radial profile of a galaxy's rotational speed.

Another `anomaly' of a low-resolution part of the universe would appear as a pressure in a continuous physical model.  Suppose an evacuated volume of spacetime were surrounded by a gas.  If the evacuated volume were instead comprised of relatively few pixels, the entropic origin of its supposed low pressure would be nullified; after all, there would be few lattice vertices for the surrounding gas to occupy.  Similarly, the expansion of a single lattice pixel via the evolution of $e^\mu_{~a}\n$ would seemingly require less energy than the expansion of an analogous volume of spacetime.

These observations invite the appealing speculation that the supposed effects of dark matter and dark energy could be reinterpreted as the tell-tales of a discrete universe.  The uncountability of spacetime in the `volumes' of lattice hypercubes might lead in continuum models to the residual matter and energy that have been estimated to comprise $\sim95\%$ of the known universe \cite{jarosik_seven-year_2011}.  We also note that, if desired, a cosmological constant can be introduced in a discrete EC gravity formalism via the vierbeins, as in \cite{smolin_quantum_1979}.

We may further consider 5-vector theory from a quantum theoretical point of view.  We note that path integral \cite{feynman_space-time_1948} formulations of quantum field theories are better-defined on a lattice than in continuous spacetime.  Therefore, we hope that our discrete, Poincar\'e-symmetric theory, once quantized, might admit unitary time evolution.  We may also consider fluctuations in $e_\mu\n$---the translation components of the solder field---from a quantum mechanical perspective.  Examining the expression
\begin{eqn}
\phi&=\phi_0-e_\mu\phi^\mu\\
&=\phi_0-e_\mu\eta^{\mu\sigma}e_\sigma^{~b}\md_b^+\phi_0,
\end{eqn}
one notes that quantum fluctuations in $e_\mu$ might appear as fluctuations in the `location' of the field $\phi\n$.  In a sense that only a fully quantized theory would make rigorous, longitudinal gravitational fluctuations in 5-vector theory appear to give rise to an effect reminiscent of Heisenberg uncertainty.

We recall that the translation components $e_\mu$ remained indeterminate in the EOM of Eq.~(\ref{SolderGaugeEOM}), and that we omitted the translation DOF from our definition of gauge curvature in Eq.~(\ref{discreteCurvature}).  It should be noted that, when simulating 5-vector theory classically, these DOF pose no obstacle; they can be set to an arbitrary constant value---e.g., zero---without penalty.  Should longitudinal gravity waves ever be observed, it would be natural to extend our theory to describe a `translation curvature'.  We do not attempt to understand the apparent absence of such a force, but we note that prior work on the role of symmetry breaking in gravity---as described in \cite{wise_geometric_2012}, for example---appears to offer an attractive framework for exploring this matter.

In some ways, lattice 5-vector theory seems more natural than previous lattice gauge theories of gravity.  In the main, such theories have focused on spinorial representations of the Lorentz gauge field, presumably due to a desire to produce a theory that couples to the fermionic matter of the Standard Model.  However, it is not clear how a discrete spinorial gauge theory of gravity would couple to a scalar particle like the Higgs boson.  Reinterpreting the scalar Higgs field as a 5-vector, on the other hand, naturally prescribes such a coupling.

Because of the matter-coupling defined in gauged 5-vector theory, we have been able to employ a Poincar\'e gauge field whose Lorentz components are in the standard representation.  Cartan's frame field therefore finds natural purchase in the solder field of 5-vector theory, and serves as a map between the standard Lorentz gauge field and the affine connection, as in Eq.~(\ref{NewGaugeField}).  Indeed, the standard Lorentz representation, embodied in the index $\mu$ of $e^\mu_{~a}\n$, comports with our notion of the metric ${g_{ab}=\eta_{\mu\nu}e^\mu_{~a}e^\nu_{~b}}$ as measuring distances in a pseudo-Riemannian spacetime.

In this sense, we may speculate as to the finiteness of a quantized version of our lattice theory of gravity.  On the one hand, the lattice has countably many discrete data points.  On the other hand, our solder field may nonetheless gauge transform to an infinitesimal UV scale.  This duality of discrete and continuous scales requires further study.

Although we will not pursue them here in any detail, fermionic extensions of the 5-vector are apparently defined without obvious additional complication.  Following the technique of 5-vector theory, the spacetime derivatives ($\partial_\mu\psi$) of a Dirac fermion would be `internalized' within $\psi$---say, as $\psi^\mu$.  For example, a 20-vector $\gv{\psi}$ could incorporate the four components of the Dirac fermion and the 16 spacetime derivatives of these components.  Such a higher dimensional Poincar\'e representation would naturally incorporate both spinorial and standard Lorentz transformations in a single field.

One prediction of 5-vector theory is that seemingly scalar particles---such as the Higgs boson---have distinct antiparticles of identical mass.  Thus, there would exist a twisted 5-vector that would presumably correspond to the negative frequency modes of the scalar Higgs field.

A more complete set of gauge particles could also be added to 5-vector theory without any obvious complication.  This would involve expanding the structure group of the discrete principal Poincar\'e bundle.  For example, a discrete universe with the four fundamental forces might be described by the following discrete principal bundle:
\begin{eqn}
P\defeq\{\v{n}\}\times G^+\times SU(3)\times SU(2)\times U(1).
\end{eqn}

\subsection{Mathematical Considerations}

\subsubsection{A noncanonical gauge theory}

Throughout our construction of gauged 5-vector theory, we have attempted to highlight its differences from a canonical Yang-Mills gauge theory.  While the presence of the solder field in our definition of gauge curvature is already striking, the most important difference appears to be the already-local symmetry of the ungauged Lagrangian.    The very purpose of gauge fields is modified in our theory; rather than restore local symmetry to the Lagrangian, the gauge fields serve to restore local symmetry to conservation laws.

We furthermore note that the conservation laws of our theory, expressed in Eqs.~(\ref{consLawFiveVectorMatrixTheory}) and (\ref{consLawGaugedFiveVectorMatrixTheory1}), were not derived from a Noether procedure; we intuited them, primarily from the development of continuous 5-vector theory in our companion paper.  We hope that a mathematical formalism that accommodates the noncanonical features of our gauge theory would suggest its own Noether-like procedure---and thereby more systematically generate the conservation laws we have identified.

\subsubsection{A discrete gauge-theoretic variational calculus}

We have employed a few elements of novel mathematical machinery in this paper---most crucially, the discrete covariant Euler operator $\mathring{\ws{E}}_q$ defined in Appendix B.  $\mathring{\ws{E}}_q$ essentially defines a new variational derivative for discrete gauge theories.  However, we do not systematically develop a gauge covariant lattice variational calculus.  Such a mathematical effort---which might extend the tools of \cite{hydon_variational_2004} and \cite{bleecker_textbook_1981}, for example, to a gauged lattice---might substantially deepen the results we have derived here.

\section{Appendix A}
We include an appendix to outline the tools of discrete variational calculus necessary to pursue the canonical Noether procedure in a discrete ungauged theory.  An introduction of notation is helpful here.  For a system of $M$ independent variables $\{x^i\}$ and $N$ dependent variables $\left\{u^\ell\right\}$, we denote a \emph{$(k\geq0)$-order multi-index} $\v{J}$ by $\v{J}\equiv(j_1,\dots,j_k)$, where ${1\leq j_i\leq M}$.  We let $\#\v{J}$ denote the order (i.e., length) of the multi-index $\v{J}$, where any repetitions of indices are to be double-counted.  We let ${\left(\begin{smallmatrix}\v{I}\\\v{J}\end{smallmatrix}\right)\equiv \v{I}!/[\v{J}!(\v{I}\backslash\v{J})!]}$ when ${\v{J}\subseteq\v{I}}$, and $0$ otherwise.  We define $\v{I}!=\left(\tilde{i}_1!\cdots\tilde{i}_M!\right)$, where $\tilde{i}_a$ denotes the number of occurrences of the integer $a$ in multi-index $\v{I}$.  $\v{I}\backslash\v{J}$ denotes the set-difference of multi-indices, with repeated indices treated as distinct elements of the set.

Let us define the higher Euler operators as in \cite{hydon_variational_2004}:
\begin{eqn}
\ws{E}_{u^\ell}^\v{J}(\ms{L}\n)\defeq\sum\limits_{\v{I}\supseteq\v{J}}\left(\begin{matrix}\v{I}\\\v{J}\end{matrix}\right)S^{-\v{I}}\pd{\ms{L}\n}{u^\ell[\v{n}+\v{I}]}.
\end{eqn}
For a vertical variational symmetry $\v{v}$ of a Lagrangian $\ms{L}\n$, we may derive its canonical conservation law $\ws{Div}A=0$ from the total homotopy operator of \cite{hydon_variational_2004}.  For $\v{v}$ defined by its characteristics $Q^\ell$ and satisfying ${\ws{pr}[\v{v}](\ms{L}\n)=0}$, the $M$-tuple $A$ is determined by the higher Euler operators, as follows:
\begin{eqn}
A^a=\sum\limits_{\ell=1}^N\sum\limits_{\#\v{I}\geq0}\frac{\tilde{i}_a+1}{\#\v{I}+1}(S-\mOne)^\v{I}\left(Q^\ell\ws{E}_{u^\ell}^{\v{I},a}(\ms{L}\n)\right).
\end{eqn}

To systematically discover the conservation laws of 5-vector theory via the discrete Noether procedure, the ``least" index that appears in the Lagrangian $\ms{L}\n$ should be ${\geq\v{n}}$.  For our present purposes, since $\md_b^-$ appears in Eq.~(\ref{theAction}), this requires that we shift the `base point' of our Lagrangian forward.  One may correspondingly denote
\begin{eqn}
\v{i}&\defeq\sum\limits_a\hat{a}=\hat{t}+\hat{x}+\hat{y}+\hat{z}
\end{eqn}
and define the \emph{shifted Lagrangian} $\ms{L}_\v{i}$ as follows:
\begin{eqn}
\ms{L}_\v{i}\n\defeq&\gv{\tphi}\nihat\v{e}^T\nihat\gv{\ws{d}}_m\v{e}\nihat\gv{\phi}\nihat\\
=&\tphi^\mu\nihat g_{\mu\nu}\nihat\phi^\nu\nihat\\
&\hspace{10pt}-\tphi^\mu\nihat e_\mu^{~a}\nihat\md_a^+\phi_0\nihat\\
&\hspace{10pt}-\tphi_0\nihat\md_b^-(e_\nu^{~b}\nihat\phi^\nu\nihat)\\
&\hspace{10pt}+m^2\tphi_0\nihat\phi_0\nihat.
\end{eqn}
The Noether procedure is then performed on $\ms{L}_\v{i}\n$, as in Eq.~(\ref{trivialDiscreteTranslationCurrent}).

\section{Appendix B}
In this appendix, we define the \emph{discrete gauge covariant Euler operator} $\mathring{\ws{E}}_q$ used to derive our gauged EOM in Eq.~(\ref{GaugedFiveVectorEOM}).  We first recall the (non-covariant) discrete Euler operator defined in Eq.~(\ref{eulerOperator}):
\begin{eqn}
\ws{E}_q(\ms{L}\n)\defeq&\sum\limits_\v{m}S^{-\v{m}}\pd{\ms{L}\n}{q[\v{n}+\v{m}]}\\
=&\sum\limits_\v{m}\pd{\big(S^{-\v{m}}\ms{L}\n\big)}{q[\v{n}]}\\
=&\sum\limits_\v{m}\pd{\big(S^{\v{m}}\ms{L}\n\big)}{q[\v{n}]}.
\label{eulerOperatorInternalized}
\end{eqn}
In the second equality, we have noted that the shift operator can be brought within $\partial/\partial q[\v{n}+\v{m}]$ so long as $q$ is shifted accordingly.  In the last equality, and simply out of preference, we have used the sum over all $\v{m}$ to flip the sign of the shift operator.

The covariant Euler operator $\mathring{\ws{E}}_q$ is now straightforwardly generalized from Eq.~(\ref{eulerOperatorInternalized}) as follows:
\begin{eqn}
\mathring{\ws{E}}_q(\ms{L}\n)\defeq&\sum\limits_\v{m}\pd{\big(\mathring{S}^\v{m}\ms{L}\n\big)}{q[\v{n}]}\\
=&\sum\limits_\v{m}\pd{\big(e^{-\v{m}}S^{\v{m}}\ms{L}\n\big)}{q[\v{n}]}.
\label{discreteGaugeCovariantEulerOperator}
\end{eqn}
This definition employs the notations $\mathring{S}^\v{m}$ and $e^{-\v{m}}$ used to define the discrete covariant derivative in Eq.~(\ref{covariantDerivative}).  Intuitively, ${\mathring{\ws{E}}_q}$ shifts the Lagrangian and then parallel transports it back to $\v{n}$ for differentiation.  Since
\begin{eqn}
\ms{S}\defeq\sum_\v{n}\ms{L}\n=\sum_\v{m}S^\v{m}\ms{L}\n,
\end{eqn}
$\mathring{\ws{E}}_q$ can be viewed as parallel transporting each term of the action $\ms{S}$ back to $\v{n}$.

A rigorous demonstration of the viability of the operator $\mathring{\ws{E}}_q$ would require a robust discussion of the variational complex on the discrete principal bundle---extending the variational complex of \cite{hydon_variational_2004}, for example---which falls beyond the scope of our present efforts.

However, we observe that $\mathring{\ws{E}}_q$ satisfies a minimal requirement: ${\ms{im}~\mD_a^\pm\subset\ms{ker}~\mathring{\ws{E}}_q}$.  That is,
\begin{eqn}
\Big[\mathring{\ws{E}}_q\circ\mD_a^\pm\Big]Q(\v{n},q_i)=0
\end{eqn}
for any function $Q$ of the lattice coordinates $\{\v{n}\}$ and dependent fields $\{q_i\}\ni q$.

\section{Acknowledgments}
Thank you to Professor Nathaniel Fisch for his early encouragement of this effort.  Thank you to Yuan Shi, whose intuitions helped spur this work.  And thank you to Sebastian Meuren, for several helpful discussions.  This research was supported by the U.S. Department of Energy (DE-AC02-09CH11466).

\end{document}